\documentclass[12pt]{article}
\usepackage{epsfig,amsmath,graphics}
\usepackage{graphics,graphicx}
\usepackage{tabularx}
\usepackage{amsfonts}
\usepackage{amsmath}
\usepackage[usenames]{color}
\usepackage{epstopdf}
\usepackage{verbatim}

\usepackage{fancyhdr}            % NB: E=pagina pari, O=pagina dispari, H=intestazione, F=piede pagina, L,R,C=sinistra,destra,centro ; \leftmark = capitolo ; \rightmark = paragrafo
\fancypagestyle{plain}{%
  \fancyhead{}                                     % Pulisce il campo della testata
  \fancyfoot[CE,CO]{\thepage}       % Numero pagina in basso ed esterno in grassetto
  \fancyhead[CE,CO]{\textcolor[rgb]{0.64,0.15,0.15}{Published in \emph{International Journal of Solids and Structures, 78-79 (2016), 205-215}
\\
doi: 10.1016/j.ijsolstr.2015.06.004}}
\setlength{\headheight}{22.5pt}}
\def\ds{\displaystyle}

\newcommand{\beq}{\begin{equation}}
\newcommand{\eeq}{\end{equation}}
\newcommand{\lb}{\label}
\newcommand{\beqar}{\begin{eqnarray}}
\newcommand{\eeqar}{\end{eqnarray}}
\newcommand{\barr}{\begin{array}}
\newcommand{\earr}{\end{array}}

\renewcommand{\d}{\textrm{d}}

\def\XXint#1#2#3{{\setbox0=\hbox{$#1{#2#3}{\int}$}
     \vcenter{\hbox{$#2#3$}}\kern-.5\wd0}}

\def\bC{\mbox{\boldmath${\it C}$}}

\def\bD{\mbox{\boldmath${\it D}$}}

\def\bE{\mbox{\boldmath${\it E}$}}
\def\be{\mbox{\boldmath${\it e}$}}
\def\bF{\mbox{\boldmath${\it F}$}}

\def\bI{\mbox{\boldmath${\it I}$}}

\def\bM{\mbox{\boldmath${\it M}$}}

\def\bN{\mbox{\boldmath${\it N}$}}

\def\b0{\mbox{\boldmath${\it 0}$}}

\def\bP{\mbox{\boldmath${\it P}$}}

\def\bS{\mbox{\boldmath${\it S}$}}

\def\bX{\mbox{\boldmath${\it X}$}}
\def\bx{\mbox{\boldmath${\it x}$}}

\def\bvarphi{\mbox{\boldmath${\varphi}$}}
\def\bsigma{\mbox{\boldmath${\sigma}$}}
\def\btau{\mbox{\boldmath${\tau}$}}

\def\tr{{\rm tr}}

\def\div{{\rm div}}
\def\Div{{\rm Div}}

\def\curl{{\rm curl}}
\def\Curl{{\rm Curl}}
\def\Grad{{\rm Grad}}
\def\grad{{\rm grad}}

\def\APL{{Appl. Phys. Lett.\ }}

\def\CMAME {{ Comput. Meth. Appl. Mech. Engrg.\ }}

\def\EJMA{{ Eur.~J.~Mechanics-A/Solids\ }}

\def\IJSS{{ Int. J. Solids Structures\ }}

\def\JAP{{ J. Appl. Phys.\ }}

\def\JMPS{{ J. Mech. Phys. Solids\ }}

\def\SMS{{ Smart Mater. Struct.\ }}

\def\salto#1#2{
[\mbox{\hspace{-#1em}}[#2]\mbox{\hspace{-#1em}}]}

\newcommand{\singlespace} {\baselineskip=12pt
\lineskiplimit=0pt \lineskip=0pt }
\def\ds{\displaystyle}

\begin{document}

\title {Analysis of viscoelastic soft dielectric elastomer generators operating in an electrical circuit}

\author{ E. Bortot$^{a}$,  R. Denzer$^{b}$\footnote{Corresponding author. {\it E-mail address}: ralf.denzer@solid.lth.se}, A. Menzel$^{c,b}$,
M. Gei$^{d}$\\
\\
\small{$^a$ Department of Civil, Environmental and Mechanical Engineering,}\\
\small{University of Trento, via Mesiano 77, I-38123 Trento, Italy}\\
\small{$^b$ Division of Solid Mechanics, Lund University, P.O.~Box 118, SE-22100 Lund, Sweden}\\
\small{$^c$ TU Dortmund University, Leonhard-Euler-Str.~5, D-44227 Dortmund, Germany}\\
\small{$^d$ School of Engineering, Cardiff University, Cardiff CF24 3AA, Wales, UK}
}
\date{}
\maketitle

\thispagestyle{plain}

\section*{Abstract}

A predicting model for soft Dielectric Elastomer Generators (DEGs) must consider a realistic model of the electromechanical
behaviour of the elastomer filling, the variable capacitor and of the electrical circuit connecting all elements of the device.
In this paper such an objective is achieved by proposing a complete framework for reliable simulations of soft energy harvesters.
In particular, a simple electrical circuit is realised by connecting the capacitor, stretched periodically by a source of mechanical work, in parallel with a battery through a diode and with an electrical load consuming the energy produced. The electrical model comprises resistances simulating the effect of the electrodes and of the conductivity current invariably present through the dielectric film.
As these devices undergo a high number of electro-mechanical loading cycles at large deformation, the time-dependent response of the material must be taken into account as it strongly affects the generator outcome. To this end, the viscoelastic behaviour of the polymer and the possible change of permittivity with strains are analysed carefully by means of a  proposed coupled electro-viscoelastic constitutive model, calibrated on experimental data available in the literature for an incompressible polyacrylate elastomer (3M VHB4910). Numerical results showing the importance of time-dependent behaviour on the evaluation of performance of DEGs for different loading conditions, namely equi-biaxial and uniaxial, are reported in the final section.

%\linenumbers

\section{Introduction}

\singlespace
In recent years the problem of energy efficiency has become more and more relevant and many efforts have been made in order to develop devices that are able to harvest energy from renewable resources.
Among the various energy harvesting technologies, Dielectric Elastomer Generators (DEGs), or dielectric elastomer energy harvesters, are particularly promising \cite{AndersonReview,antoniadis_sms2013,SRI2011,kornbluhpelrine_etc2011,AndersonMcKay,vertechy_spie2013,vertechy_spie2014,KaltseisKeplingerKohEtAl2014}.
A DEG is an electromechanical transducer based on the high deformations achievable by a filled parallel-plate capacitor subject to a voltage, constituted of a soft dielectric elastomer film usually made up of acrylic or natural rubber embedded between two compliant electrodes. By performing an electromechanical cycle in which the system is excited by an external mechanical source from a contracted to a stretched configuration at different voltages, it is possible to harvest a net energy surplus.
Evaluation of the potential amount of energy that can be harvested by a DEG in a cycle ranges between a few tens to a few hundreds of mJ/g \cite{noi_cer2014,HuangSuo,KaltseisKeplingerBaumgartnerSuo_baloon,KaltseisKeplingerKohEtAl2014,AndersonMcKay,springh_IMAjournal}.

When the generator operates effectively in a natural energy harvesting field, it will undergo a high number of electromechanical cycles at frequencies ranging from a few tenths of Hz to a few Hz and at quite high stretches. Hence, on the one hand, time-dependent effects such as viscosity of the elastomer \cite{AskMenzelRistinmaa1,AskMenzelRistinmaa2,hong,wang_visco2013} may considerably modify the performance of the generator and for this reason cannot be neglected.
On the other hand, the high strains involved in the membrane justify the analysis with electrostriction, i.e.\ the dependency of the dielectric permittivity on the mechanical stretch, even though this phenomenon depends
on the analysed material and its measurement may be strongly conditioned by the testing conditions \cite{Tagarielli,wisslermazza2007,zhaosuo2008,McKay_dielectricconstant,dilillo_mazza2012,gal_cohen2014}.

Some recent papers are devoted to the analysis of the performance of dielectric elastomer generators and, among these papers, a few take the presence of dissipative effects into consideration.
By neglecting dissipation, in \cite{KohSuo1} and \cite{springh_IMAjournal} the performance of the generator is analysed and optimised with respect to the typical failure modes of the dielectric elastomer.
%Furthermore \cite{KaltseisKeplingerBaumgartnerSuo_baloon} introduces an experimental method, based on a balloon-like generator, which tracks both electrical and mechanical energy flow during the harvesting cycle.
In \cite{FooKohKeplinger}, \cite{HuangSuo} and \cite{vertechy_spie2013}, the analysis of the performance of a dissipative dielectric elastomer generator is presented. Whereas in \cite{FooKohKeplinger} and \cite{vertechy_spie2013} the dielectric membrane and the external circuits are coupled by means of electromechanical switches, in \cite{HuangSuo} the generator is integrated in an electrical circuit constantly supplied by a battery. This simple kind of harvesting circuit with constant power supply is used in several experimental studies and is considered in \cite{pelrine_prahlad} and \cite{AndersonReview}.
M\"unch et al.\ \cite{Munch} describe the coupling of a ferroelectric generator and an electric circuit in order to determine the working points of the device. Sarban et al.\ \cite{SarbanLasseWillatzen} develop an analysis for a dielectric elastomer actuator based on the coupling of an electric circuit with a viscoelastic mechanical model.

The present paper has several objectives. First, we aim at proposing the  analysis of a soft energy harvester connected to an electric circuit
where a battery at constant voltage supplies the charge at low electric potential {\bf and electric field} to the generator{\bf, thus avoiding the electric breakdown and limiting the leakage dissipation}. Resistance of electrodes and conductivity of the dielectric are taken into account {\bf according to ohmic modelling of the leakage current}.
Secondly, we take into account the pronounced viscoelastic and electrostrictive behaviour of the material at large strains. The third objective is the analysis of such a system under typical operating conditions.
{\bf In the investigation, inertia effects are disregarded as the kinetic energy computed along the imposed oscillations is negligible with respect to the elastic strain energy stored in the elastomer}.

The paper is organised as follows.
In section 2, we will start presenting the electrical circuit for energy harvesting, in which the generator operates. This leads us to a set of nonlinear differential algebraic equations.
Then, in section 3, we will introduce a large-strain electro-viscoelastic model of the elastomer, following the approach proposed by Ask et al.\ in \cite{AskMenzelRistinmaa1,AskMenzelRistinmaa2}. Moreover, we will introduce a model for electrostriction, referring to that proposed by Gei et al.\ in \cite{GeiColonnelliSpring2014}. The model will be validated on the basis of experimental data reported in \cite{Tagarielli} for an acrylate elastomer VHB-4910 produced by 3M.
Finally, in section 4, we will present and compare the numerical results obtained for different loading conditions, i.e.  equi-biaxial and uniaxial load, and for different constitutive models, i.e. a hyperelastic solid, a viscoelastic and an electrostrictive viscoelastic material.

\section{Dielectric elastomer generator: electric circuit}
\label{sec_elec}

We consider a soft dielectric generator consisting of a block of thin soft dielectric elastomer with dimensions $L_0 \times L_0 \times H_0$
in the reference configuration $\mathcal{B}_0$. The device is assumed to deform homogeneously and is loaded by in-plane external oscillating forces represented by the nominal stress components $S_1(t)$ and $S_2(t)$
as depicted in Fig.~\ref{CircuitDEG}.a. The two opposite surfaces are treated so as to act like compliant electrodes inducing, neglecting fringing effects, a nominal time-dependent electric field $E^0(t)$ directed along the coordinate $X_3$.
Related to the deformation history the dimensions of the elastomer vary as a function of the time-dependent principal stretches $\lambda_i(t)$, with $i = 1,2,3$, to reach, at a certain time $t$, the actual dimensions $L_1=L_0 \,\lambda_1(t)$, $L_2=L_0\,\lambda_2(t)$ and $H=H_0\,  \lambda_3(t)$.

This generator can generally be modelled as a stretch-dependent variable plane capacitor, the capacitance $C$ of which is defined as
\beq\lb{capacitance}
C(t)=\epsilon \, \frac{A}{H} = \epsilon \, \frac{L_0^2}{H_0} \, \frac{\lambda_1(t) \lambda_2(t)}{\lambda_3(t)} \, ,
\eeq
where $\epsilon$ is the dielectric permittivity that can be decomposed as $\epsilon=\epsilon_r \,\epsilon_o$. Moreover, $\epsilon_r$ represents the relative dielectric constant and {\bf $\epsilon_o = 8.854$~pF/m} characterises the permittivity of vacuum.

In a real device, however, the dielectric material shows a certain conducting current, also denoted as leakage current, while the electrodes have a non-negligible resistance. Hence, a more realistic electrical model of the generator is a variable capacitor connected in parallel to a resistor $R_i$, representing the electrical resistance of the dielectric film, and connected in series to a resistor $R_s$, representing the electrical resistance of electrodes and wires, as shown in Fig.~\ref{CircuitDEG}.b, see \cite{SarbanLasseWillatzen}.

Furthermore, the charge $Q$ exchanged by the system is given by the sum of the time-integral of the leakage current and the product of capacitance and voltage of the soft variable capacitor,
\beq\lb{CapacitorCharge}
Q(t) = \int_{0}^{t} i_{Ri}(\tau) \,d \tau + C(t) \phi_{_C}(t) \, .
\eeq

%%%%%%%%%%%%%%%%%%%%%%%%%%%%%%%%%%%%%%%%%%%%%%%%%%%%%%%%%%%%%%%%%%%%%%
\begin{figure}[!t]
  \begin{center}
\includegraphics[width= 6 cm]{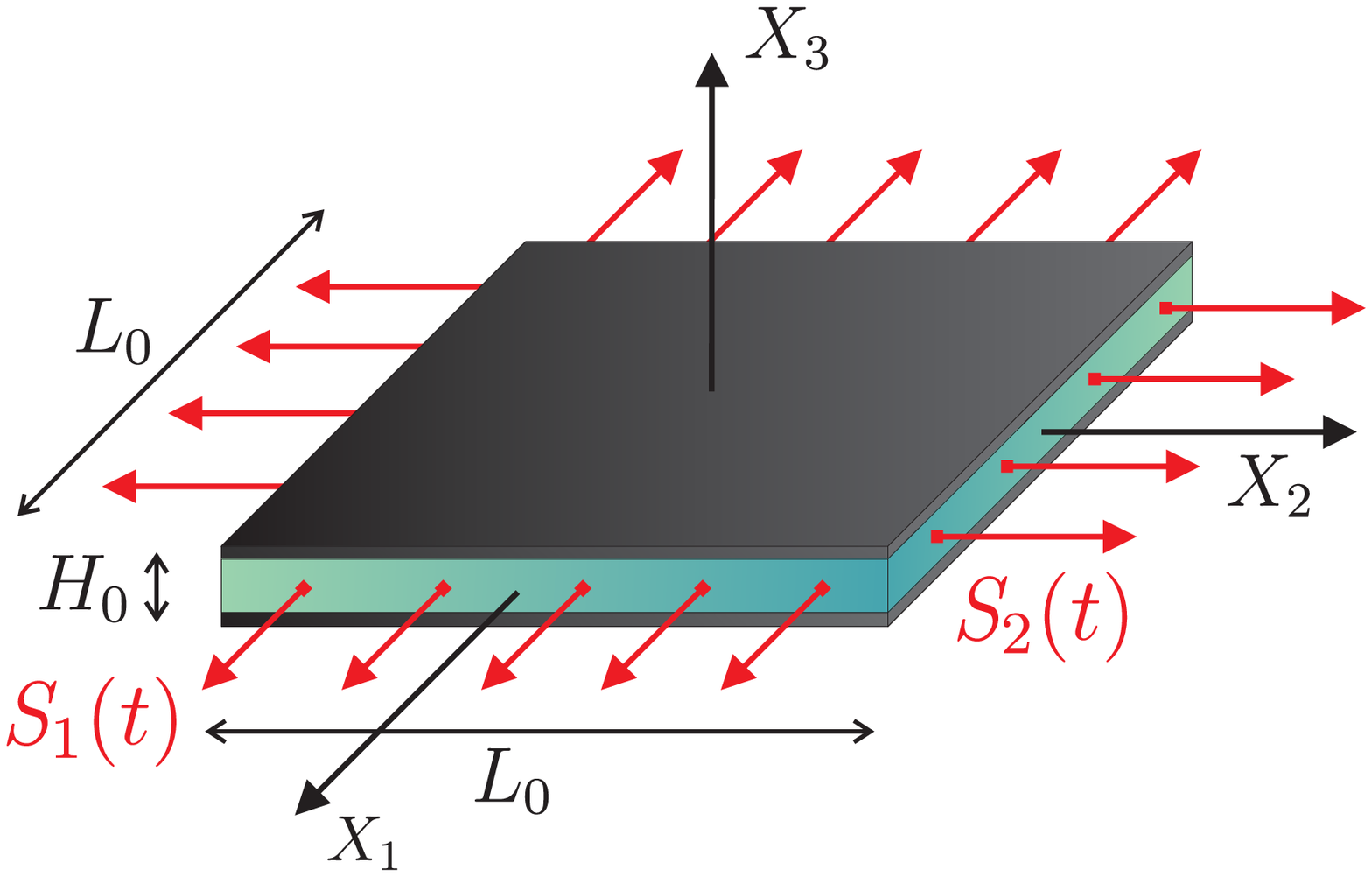}~a)
   \qquad
\includegraphics[width= 7 cm]{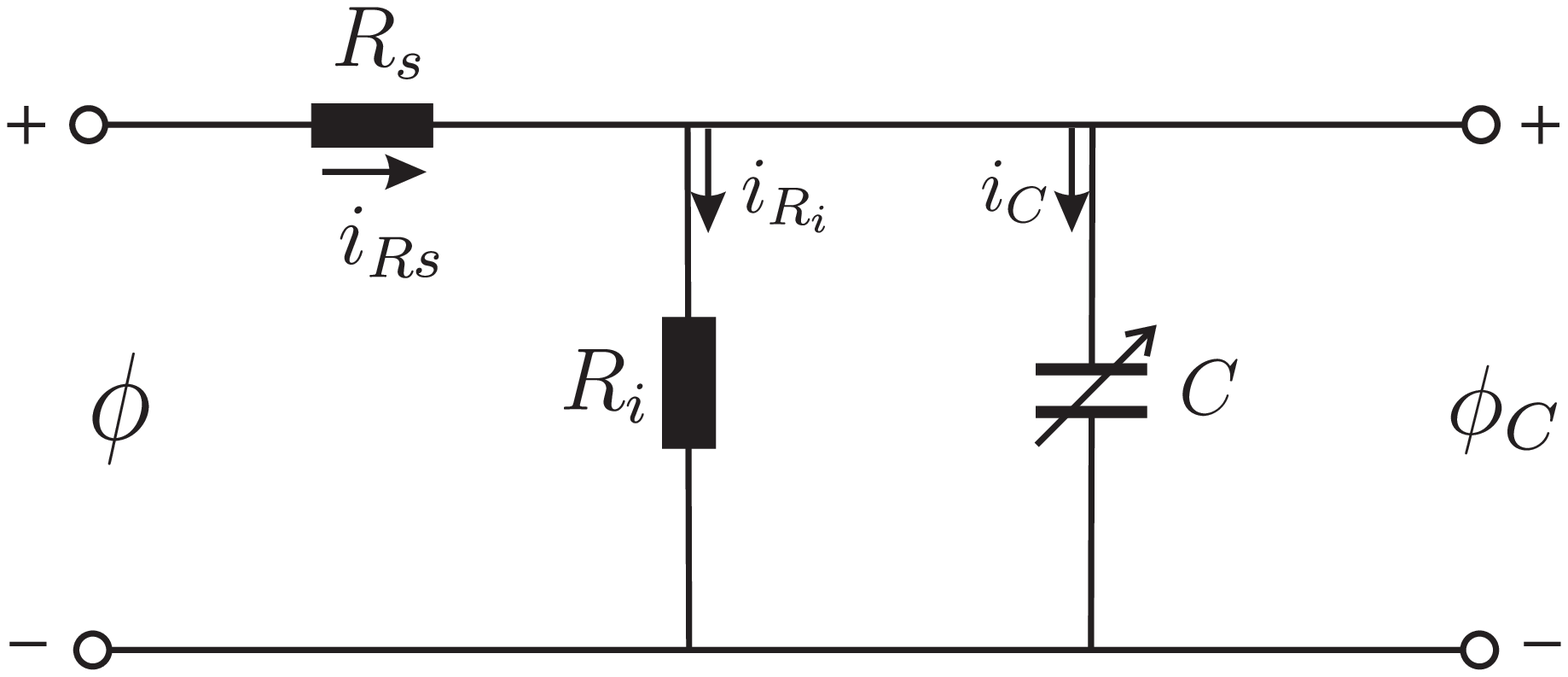}~b)
\vspace{0.5 cm} \caption{{a) Dielectric elastomer generator in its reference configuration; b) scheme of the equivalent circuit diagram of a soft dielectric elastomer generator.}}
    \lb{CircuitDEG}
 \end{center}
\end{figure}
%%%%%%%%%%%%%%%%%%%%%%%%%%%%%%%%%%%%%%%%%%%%%%%%%%%%%%%%%%%%%%%%%%%%%%

The generator operates in an electrical circuit achieved by connecting the dielectric elastomer generator in parallel to a battery through a diode and to an electrical load, as illustrated in Fig.~\ref{Circuit}.
The battery supplies the circuit with a difference in the electric potential $\phi_o(t)$. In the analysis of the circuit, we assume that the voltage supplied by the battery is zero at the initial time $t=0$ and then increases linearly during the semi-period $T/2$ of the stretch oscillation up to the value $\phi_o$, namely
\beq
\lb{ciccio}
\phi_o(t)= t \,\frac{\phi_o}{T/2}  \quad \mbox{for} \quad 0<t<T/2 \, .
\eeq
Thereafter, for $t>T/2$, the supplied voltage is kept constant, i.e.
$$
\phi_o(t)=\phi_o  \quad \mbox{for} \quad t>T/2\, .
$$

%%%%%%%%%%%%%%%%%%%%%%%%%%%%%%%%%%%%%%%%%%%%%%%%%%%%%%%%%%%%%%%%%%%%%%
\begin{figure}
  \begin{center}
\includegraphics[width= 11 cm]{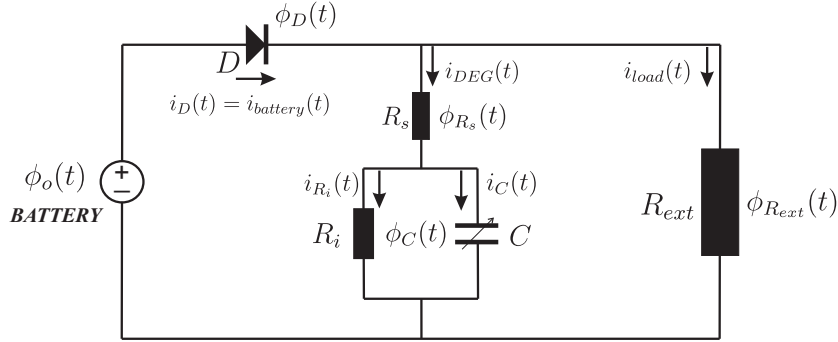}
\vspace{0.5 cm} \caption{{Scheme of the electrical circuit in which the dielectric elastomer generator operates.}}
    \lb{Circuit}
 \end{center}
\end{figure}
%%%%%%%%%%%%%%%%%%%%%%%%%%%%%%%%%%%%%%%%%%%%%%%%%%%%%%%%%%%%%%%%%%%%%%

The electrical load is represented by the external resistor $R_{ext}$. The impedance of the load has to be sufficiently high so that the charge is maintained constant during the release of the elastomer and, as a consequence, the voltage on the dielectric elastomer is increased with respect to the constant value $\phi_o$ supplied by the battery.

The diode prevents the charge from flowing from the generator to the battery during the release phase. Its current $ i_{_{D}}(t)$ is modelled according to the classical Shockley diode equation
\beq\lb{shockleyDiode}
i_{_{D}}(t)=I_s\left[\exp {\left(\frac{\phi_{_{D}}(t)}{n\; v_T}\right)}-1\right],
\eeq
where $I_s$ is its saturation current, $v_T$ the thermal voltage, $n$ the ideality factor with $1<n<2$, and $ \phi_{_{D}}(t)$ the diode voltage.
The thermal voltage depends on the Boltzmann constant $K$, the temperature $T$ and on the elementary charge $q_e$ = 1.60217653 x $10^{-19}$~C, as $v_T=KT/q_e$.

In the case where the components of a circuit are connected in series, the total voltage is equal to the sum of the voltage on each of the components. By applying Kirchhoff's voltage law to the circuit one obtains
\beqar
\lb{circuitEq.voltage}
\phi_o(t)&=&\phi_{_{D}}(t)+\phi_{_{Rs}}(t)+\phi_{_{C}}(t)\, ,\\
\lb{circuitEq.voltage2}
\phi_o(t)&=&\phi_{_{D}}(t)+\phi_{_{Rext}}(t)\, ,
\eeqar
where $\phi_{_{C}}(t)$ is the voltage on the generator and the parallel resistor $R_i$, while $\phi_{_{Rext}}(t)$ is the voltage on the electric load, here represented by the external resistor with impedance $R_{ext}$.
Combining (\ref{circuitEq.voltage}) and (\ref{circuitEq.voltage2}) results in the voltage relation for a parallel connection,
$$
\phi_o(t)-\phi_{_{D}}(t)=\phi_{_{Rs}}(t)+\phi_{_{C}}(t)=\phi_{_{Rext}}(t)\, .
$$
Recalling that series-connected circuit elements carry the same current while parallel-connected circuit elements share the same voltage, so that the overall current is the sum of the currents on each element, we can describe the circuit by using Kirchhoff's current law
\beq
\lb{circuitEq.current}
i_{_{D}}(t)=i_{_{battery}}(t)=i_{_{DEG}}(t)+i_{_{load}}(t) \, .
\eeq
{\bf Experiments on acrylic elastomers \cite{dilillo_mazza2012} have shown that the response of resistors $R_i$ and $R_s$ is ohmic if the electric field in the material will not exceed a threshold value in the range between 20 and 40 MV/m, beyond which the resistance will decrease exponentially.
In our simulations we take the voltage $\phi_o$ supplied by the battery at constant regime as 1 kV and therefore the intensity of the electric field in the generator remains bounded to 20 MV/m. As a consequence, we assume
Ohm's laws
$i_{_{DEG}}(t)=\phi_{_{Rs}}(t)/R_s$ and $i_{_{Ri}}(t)=\phi_{_{C}}(t)/R_i$ to complete the formulation.
}
Therefore, eq.~(\ref{circuitEq.current}) together  with (\ref{circuitEq.voltage}) and (\ref{circuitEq.voltage2}) constitute a non-linear differential algebraic system of four equations
\beq\lb{systemCircuit}
   \begin{cases}
\ds{    \phi_o(t)-\phi_{_{D}}(t)=\phi_{_{Rs}}(t)+\phi_{_{C}}(t)\, ,} \vspace{4mm} \\
\ds{     \phi_{_{Rs}}(t)+\phi_{_{C}}(t)=\phi_{_{Rext}}(t) \, ,} \vspace{4mm} \\
\ds{     I_s\left[\exp\left(\frac{\phi_{_{D}}(t)}{n \; v_T}\right)-1\right] =  \frac{\phi_{_{Rs}}(t)}{R_s}+\frac{\phi_{_{Rext}}(t)}{R_{ext}} \, ,} \vspace{4mm} \\
\ds{     \frac{\phi_{_{Rs}}(t)}{R_s} = C(\lambda(t)) \; \dot{\phi}_{_{C}}(t)+ \dot{C}(\lambda(t))\; \phi_{_{C}}(t)+\frac{\phi_{_{C}}(t)}{R_i} \, ,}
   \end{cases}
\eeq
where the voltages $\phi_{_{D}}(t)$, $\phi_{_{Rs}}(t)$, $\phi_{_{C}}(t)$ and $\phi_{_{Rext}}(t)$ are the four unknowns. The non-linear system (\ref{systemCircuit}) can be solved numerically, e.g.\ by using a DAE solver. Schuster \cite{Schuster} presents the recourse to differential algebraic equation solvers in the analysis of nonlinear electric networks.
{\bf Regarding the values of resistances in the circuit, on one hand, a review of the literature
\cite{HausDarmstadt1,HausDarmstadt2,SchlaakDarmstadt3} has led us to set $R_i= 100$ G$\Omega$ and $R_s=70$ k$\Omega$ as a reasonable choice. On the other, as we aim at comparing the behaviour of the generator for different end
users, we select a quite large range for $R_{ext}$, namely $R_{ext}\in[0.001,1000]$ G$\Omega$.}

%As reasonable values of the capacitor and the circuit resistances
%we take $R_i= 100$ G$\Omega$ and $R_s=70$ k$\Omega$; see \cite{HausDarmstadt1}, \cite{HausDarmstadt2} and  \cite{SchlaakDarmstadt3}.
%Moreover, we assume that the voltage $\phi_o$ supplied by the battery at constant regime is 1 kV.

For the description of the characteristic parameters of the diode, we refer to the commercial type designated as NTE517 produced by NTE Electronics Inc. In agreement with \cite{datasheetNTE517}, we estimate that the saturation current $I_s$ is $\simeq$ 0.1 $\mu$A and that the thermal voltage $v_T$ is $\simeq$ 25 mV at room temperature. In the computations, we will assume a unitary value $n = 1$ for the ideality factor of the diode.

From an electro-mechanical point of view, the soft dielectric generator consists of an incompressible electroactive polymer (EAP) to
be modelled by employing the large-strain electro-viscoelasticity framework introduced by Ask et al.\ \cite{AskMenzelRistinmaa1,AskMenzelRistinmaa2}, which is briefly summarised in the following sections. The main hypotheses lie in the assumption that the electric fields are static whereas the mechanical response, though quasi-static, is rate-dependent.

\section{Large strain electro-viscoelasticity}

\subsection{Kinematics and governing equations}
For the motion of the material body considered, we assume that $\bvarphi (\bX,t)$ is a sufficiently smooth mapping transforming the position vector $\bX$ of a material particle in the reference configuration $\mathcal{B}_0$ to its spatial position $\bx=\bvarphi (\bX,t)$ in the actual configuration $\mathcal{B}_t$ at time $t$.
Hence, the deformation gradient tensor is given by $\bF = \Grad  \bvarphi$, where the gradient is taken with respect to the reference configuration $\mathcal{B}_0$.
The local volume ratio is the Jacobian of the deformation gradient tensor $J= \det \bF$ with $J=1$ for incompressible materials.
The right Cauchy-Green tensor is defined by $\bC =\bF^T \cdot \bF$ and we formally introduce the stretches $\lambda_1$, $\lambda_2$, $\lambda_3$, already used in section \ref{sec_elec}, as the square roots of the eigenvalues of $\bC$ such that $J=\lambda_1 \, \lambda_2\,  \lambda_3 =1$.

The quantities of interest to define the electrostatic state of a dielectric are the electric field $\bE$, the electric displacements $\bD$ and the polarisation $\bP$ in $\mathcal{B}_t$, linked by the relation
$$ \bD = \epsilon_o \, \bE + \bP. $$

%In the reference configuration $\mathcal{B}_0$ these quantities are given by
%$$ \bE^0 = \bF^T \cdot \bE, \qquad \bD^0= J \bF^{-1} \cdot \bD, \qquad \bP^0= J \bF^{-1} \cdot \bP .$$
%In the continuum, $\bE$, $\bD$ and $\bP$ are related through
%In the reference configuration the previous relation turns to be
%$$ \bD^0 = \epsilon_o J \bC^{-1} \cdot \bE^0 + \bP^0. $$

\begin{sloppypar}
Electromagnetic interactions are governed by Maxwell's equations. We assume throughout the paper that i) the hypotheses of electrostatics hold true and that ii) free currents and free charges are absent. Therefore, Maxwell's equations in local form with respect to the actual configuration $\mathcal{B}_t$ reduce to
\end{sloppypar}
\beq
\curl\bE=\textbf{0} \, ,\ \ \ \
\div\bD=0\, ,
\eeq
or with respect to the reference configuration $\mathcal{B}_0$ to
\beq\lb{MaxwellEq1}
\Curl\bE^0=\textbf{0}\, ,\ \ \ \
\Div\bD^0=0\, ,
\eeq
where the following nominal fields
\beq
 \bE^0 = \bF^T \cdot \bE\, , \qquad \bD^0= J \bF^{-1} \cdot \bD \, ,
\eeq
are naturally introduced.

The notation used in eq. (\ref{MaxwellEq1}) is such that the uppercase letters indicate operators acting on $\mathcal{B}_0$, e.g.\ $\Grad,\; \Div,\; \Curl$, whereas lowercase letters refer to operators defined in the configuration $\mathcal{B}_t$, e.g.\ $\grad,\; \div, \; \curl$.
Eq.~(\ref{MaxwellEq1})$_1$ implies that the electric field is conservative, i.e.
\beq\lb{EqPotEl}
 \bE^0 (\bX)=-\Grad \phi (\bX)\, ,
\eeq
where $\phi(\bX)$ is the electrostatic potential.
At a discontinuity surface, including the boundary $\partial\mathcal{B}_0 $, the electric field and the electric displacement must fulfil the jump conditions
\beq\lb{JumpCond}
    \salto{0.1}{\bE^0} \times \bN^0={\bf 0}\, , \;  \qquad \qquad  \; \salto{0.1}{\bD^0}\cdot \bN^0=0 \, ,
\eeq
where $ \salto{0.1} {f } =  f^a- f^b$ is the jump operator and where $\bN^0$ denotes the outward referential unit normal vector, pointing from $a$ towards $b$.

The local form of the balance of linear momentum in $\mathcal{B}_t$ for the quasi static case corresponds to
\beq\lb{BalLinMom1}
\div \bsigma + \textbf{\em{f}}_e+ \rho \, \textbf{\em{f}}={\bf 0}\, ,
\eeq
where $\rho$ is the current mass density of the body, $\textbf{\em{f}}$\, is the mechanical body force and $\textbf{\em{f}}_e$ is the electric body force per unit of volume.
{\bf The inertia term is neglected as we will show that it is not substantial in the performance analysis of prestretched elastomer generators}.
For the problem at hand the electric body force can be specified as follows
$$ \textbf{\em{f}}_e = \grad \bE \cdot \bP \, .$$
Moreover, the Cauchy stress tensor $\bsigma$ is generally non-symmetric, whereas the total stress tensor
$$  \btau= \bsigma+ \bE  \otimes  \bD - \frac{1}{2} \epsilon_0 [ \bE  \cdot  \bE] \bI\, ,$$
as introduced in e.g.\ \cite{dorf&ogde05acmc,HutterVanDeVenUrsescu,Maugin,mcmeeking}, turns out to be symmetric. The second-order identity tensor is denoted by $\bI$.
In this way, it is possible to rewrite the balance of linear momentum as
$$ \div  \btau + \rho   \textbf{\em{f}} = {\bf 0} \, . $$
The total Piola-type stress tensor $\bS$ is defined as $ \bS= J \, \btau \cdot \bF^{-T}$, so that the local referential form of the balance of linear momentum can be written as
$$ \Div  \bS + \rho_0  \, \textbf{\em{f}} ={\bf 0} \,, $$
where $\rho_0 = J\rho$ is the referential mass density. In view of the inverse motion problem of electro-elasticity, respectively electro-viscoelasticity, the reader is referred to \cite{AskDenzerMenzelRistinmaa2013,DenzerMenzel2014} and references cited therein.

\subsection{Viscoelasticity at finite deformation}

The DEs are elastomers with rubber-like properties. Hence, it is relevant to extend the electro-elastic framework in order to include viscoelastic effects and to thereby model the rate-dependence mechanical behaviour of the material.
We assume that the viscosity is related to mechanical contributions only, i.e. the deformation gradient and additional internal variables which represent the viscous part of the behaviour. This means that, even though the material deforms in response of an applied electric voltage, the viscosity is related to the induced deformation only, and not directly to the electrical quantities.
In the present work, we will refer to the viscoelastic model proposed by Ask et al.\ \cite{AskMenzelRistinmaa1,AskMenzelRistinmaa2}, and to the one by Gei and collaborators \cite{BertoldiGei2011,GeiColonnelliSpring2014} for the electromechanical behaviour.

A common approach to model viscoelasticity, see e.g.\ \cite{Lubliner, ReeseGovindjee,KleuterMenzelSteinmann2007}, in the finite-strain framework is based on the introduction of a multiplicative split of the deformation gradient into elastic and viscous contributions
\beq\lb{SplitDG}
\bF = \bF_{e \alpha} \cdot \bF_{v \alpha}\, ,
\eeq
where subscript $\alpha$ indicates the possibility of multiple viscosity elements.
The multiplicative decomposition (\ref{SplitDG}) can be considered as a three-dimensional generalisation of the splitting occurring in a
one-dimensional Maxwell rheological element, where a spring and a dashpot are connected in series.
In a generalised Maxwell rheological model, an arbitrary number of Maxwell elements is connected in parallel.
For later reference, it is convenient to introduce a Cauchy-Green-type deformation tensor defined as
\beq\lb{Cvisco}
\bC_{v \alpha} = \bF_{v \alpha}^T \cdot \bF_{v \alpha}\, ,
\eeq
for each Maxwell element $\alpha$. This tensor will be taken as the internal variable and shall satisfy $\det \bC_{v \alpha}=1$.

The dissipation inequality, which is the basis to formulate constitutive equations, can be written in local form as
\beq\lb{DissipationIneq}
\mathcal{D}=\left[ \bS-\frac{\partial W}{\partial \bF} \right] : \dot{\bF} - \left[ \bD^0+\frac{\partial W}{\partial \bE^0} \right]\cdot \dot{\bE}^0-\sum_\alpha \frac{\partial W}{\partial \bC_{v\alpha}}:\dot{\bC}_{v\alpha}\geq0\, ,
\eeq
where the notation $\dot{\bullet}$ denotes the material time derivative.
The dissipation inequality must be valid for all admissible processes. Hence, a sufficient condition for the non-viscous part of (\ref{DissipationIneq}) to be fulfilled is that
\beq\lb{Non-viscousEq}
\bS=\frac{\partial W}{\partial \bF}-p\bF^{-T}\, , \qquad \qquad \bD^0=-\frac{\partial W}{\partial \bE^0}\, ,
\eeq
where $p$ is the hydrostatic pressure due to the incompressibility constraint.
In order to fully characterise the material behaviour, it is necessary to formulate evolution equations for the internal variables, which describe the rate-dependence of the mechanical quantities.

It is assumed that the elastomer is an incompressible material, so that $J=1$, complying with a constitutive relation of neo-Hookean type under isothermal conditions. Assuming the nominal electrical field $\bE^0$ as the independent electrical variable, the electric Gibbs potential is considered to take the representation
\beq\lb{WFE0}
W(\bF,\bE^0,\bC_{v\alpha})=\frac{\mu}{2}[I_1-3]+\frac{1}{2} \sum_\alpha  \beta_\alpha\, \mu \, [{I_1}_{v\alpha}-3]-\frac{\epsilon}{2} I_5\, ,
\eeq
with $I_1 = \tr \bC$, ${I_1}_{v\alpha}= \tr(\bC \cdot \bC_{v \alpha}^{-1})$ and $I_5=\bE^0 \cdot \bC^{-1}\cdot \bE^0 = \bE\cdot \bE$. Here, $\mu$ is the long-term shear modulus of the material and $\beta_{\alpha}$ are positive dimensionless proportionality factors, which relate the shear modulus of the viscous element $\alpha$ to the long-term shear modulus $\mu$.
If the dielectric permittivity $\epsilon$ is independent of the deformation, we can represent the permittivity as $\epsilon=\epsilon_0 \, \epsilon_r^0$, where $\epsilon_r^0$ is the relative permittivity referred to the undeformed configuration.
Otherwise, if the permittivity is stretch dependent, i.e.\  $\epsilon(\lambda_1, \lambda_2, \lambda_3)$, the permittivity takes the form  $\epsilon(\lambda_1, \lambda_2, \lambda_3 )=\epsilon_0 \, \epsilon_r(\lambda_1, \lambda_2, \lambda_3 )$, where ${\epsilon}_r(\lambda_1, \lambda_2, \lambda_3 )$ is the deformation dependent relative dielectric permittivity.

Based on equation (\ref{Non-viscousEq}), a necessary condition for the evolution equations of the internal variables to satisfy is
\beq
\mathcal{D}_v=-\sum_\alpha \frac{\partial W}{\partial \bC_{v \alpha}} : \dot{\bC}_{v \alpha}\geq 0\, .
\eeq
The definition of a Mandel-type referential stress tensor as
\beq\lb{MandelStress}
\bM_{v \alpha}= -   \bC_{v \alpha} \cdot \frac{\partial W}{\partial \bC_{v \alpha}}\, ,
\eeq
allows to restate the dissipation inequality in the following form
\beq\lb{ViscousDissipation}
\mathcal{D}_v=\sum_\alpha \bM_{v \alpha}:[\bC_{v \alpha}^{-1} \cdot \dot{\bC}_{v \alpha}]\geq 0\, .
\eeq
A possible format of the evolution equations which fulfills the dissipation inequality and ensures the symmetry of $\bC_{v \alpha}$, see \cite{AskMenzelRistinmaa1, AskMenzelRistinmaa2}, is given by

\beq\lb{EvolutionLaw}
\dot{\bC}_{v \alpha}= \dot{\varGamma}_\alpha \, \bC_{v \alpha} \cdot {\bM_{v \alpha}^{dev}}^T\, ,
\eeq
where $\dot{\varGamma}_\alpha$ are material parameters.

\section{Calibration of the electro-viscoelastic model}
\label{sec_calib}

The material taken into consideration is the polyacrylate dielectric elastomer VHB-4910, produced by 3M$^{\mathtt{TM}}$, assumed to show incompressible behaviour, i.e. $J=1$.
Using the energy function (\ref{WFE0}) and the constitutive equations (\ref{Non-viscousEq})$_{1,2}$, we obtain the following expressions
\begin{align}
\lb{stress}
&\bS =-p \,  \bF^{-T} + \mu\,  \bF+ \sum_\alpha \beta_\alpha \,  \mu\,   \bF \cdot \bC_{v \alpha}^{-1}+\epsilon\,   \bF^{-T}\cdot \bE^0 \otimes \bC^{-1}\cdot \bE^0\, ,\\
\lb{displacement}
&\bD^0 = \epsilon \,  \bC^{-1} \bE^0\, .
\end{align}
for the nominal stress $\bS$ and for the nominal electric displacement $\bD^0$. Furthermore, the Mandel-type referential stress tensor defined in (\ref{MandelStress}) is given by
\beq\lb{Mva}
\bM_{v \alpha}= \frac{1}{2}\,   \beta_\alpha\,   \mu\,   \bC \cdot \bC_{v \alpha}^{-1}\, ,
\eeq
so that (\ref{EvolutionLaw}) results in
\beq\lb{EvolutionLawII}
\dot{\bC}_{v \alpha}= \frac{1}{2}\,  \beta_{\alpha}\,  \mu\,  \dot{\varGamma}_\alpha  \left[ \bC -\frac{1}{3}\,  [\bC  : \bC_{v \alpha}^{-1}]\,  \bC_{v \alpha} \right]\, .
\eeq

The material parameters are identified by separating mechanical and electrical behaviour. Experimental data by Tagarielli et al.\  \cite{Tagarielli} are used for the calibration of the  electro-viscoelastic model.

\subsection{Calibration of the mechanical behaviour}

The mechanical response of the model is calibrated with experimental data based on a uniaxial tensile loading test.
In the absence of electrical effects, i.e.\ $\bE^0=  {\bf 0}$, for a uniaxial stress state -- where the Cartesian base vectors $\{\be_1, \be_2, \be_3 \}$ are assumed to coincide with the principal directions such that
$\lambda_1=\lambda(t) \, , \,\lambda_2=\lambda_3=1/  \sqrt{\lambda(t)}  $ --  the viscoelastic stress in the loading direction can be computed using (\ref{stress}),
\beq
S = \mu \, \lambda + \sum_{\alpha} \beta_{\alpha}\,   \mu\,   \frac{\lambda}{\lambda_{v  \alpha}^2}- \sum_{\alpha}\mu \, \frac{\beta_{\alpha}\,  \lambda_{v  \alpha} + 1}{\lambda^2}\, ,
\eeq
cf.\ \cite{AskMenzelRistinmaa1}. Here $\lambda_{v \alpha}$ are the internal variables formally defined as the square root of the eigenvalues of the respective $\bC_{v \alpha}= \lambda_{v  \alpha}^2\,  \be_1 \otimes \be_1 + \lambda_{v  \alpha}^{-1}\,  [\bI - \be_1 \otimes \be_1] $.

In \cite{Tagarielli} three different strain rates $\dot{\delta}_m$ are considered, namely $\dot{\delta}_1= 7 \times 10^{-3}$ s$^{-1}$, $\dot{\delta}_2=1.5 \times 10^{-2}$ s$^{-1}$ and $\dot{\delta}_3 = 3 \times 10^{-2}$ s$^{-1}$. The strain rate is held constant during the measurements, displacing the cross-head of the testing machine at a variable velocity $\dot{u}_m$ such that
\beq\lb{strainRate}
\dot{\delta}_m=\frac{\dot{u}_m}{l} =\frac{\dot{u}_m}{l_0+u_m(t)}= {\rm const}\, ,
\eeq
where $l_0$ is the initial length of the sample and where $l$ is the actual length. From equation (\ref{strainRate}), the displacement of the cross-head $u_m(t)$ can be computed by solving the ordinary differential equation $\dot{u}_m=\dot{\delta}_m\, [l_0+u_m(t)]$ under the condition $u_m(0)=0$, namely
$$
u_m(t)= l_0 \, [\exp{(\dot{\delta}_m \,  t)}-1]\, .
$$
This leads to the stretch ratio
$$
\lambda(t)=\frac{l_0+u_m(t)}{l_0}= \exp{(\dot{\delta}_m \,  t)}\, .
$$

The response of the model is compared to the experimental data obtained at discrete time points ({\em i, j, k}) for the three strain rates $\dot{\delta}_m$. The aim is to find the set of parameters $\{\mu, \beta_{\alpha},\dot{\varGamma}_{\alpha}\}$ by minimising, for all measured data points, the difference between the stress $S^{exp}$ determined experimentally and $S^{sim}$ predicted by the model.
In particular, the error to be minimised is computed  using the  $L_2$-norm as
\beq
{\rm Error}(\mu, \beta_{\alpha}, \dot{\varGamma}_{\alpha})= \sqrt{\sum_i[\Delta S_i(\dot{\delta}_1)]^2
+\sum_j[\Delta  S_j(\dot{\delta}_2)]^2+ \sum_k[\Delta  S_k(\dot{\delta}_3)]^2}\, ,
\eeq
where $\Delta S_i(\dot{\delta}_1)$, $\Delta S_j(\dot{\delta}_2)$ and $\Delta S_k(\dot{\delta}_3)$ denote the differences $[S_i^{exp}(\dot{\delta}_1)-S_i^{sim}(\dot{\delta}_1)]$, $[S_j^{exp}(\dot{\delta}_2)-S_j^{sim}(\dot{\delta}_2)]$ and $[S_k^{exp}(\dot{\delta}_3)-S_k^{sim}(\dot{\delta}_3)]$, respectively.

%%%%%%%%%%%%%%%%%%%%%%%%%%%%%%%%%%%%%%%%%%%%%%%%%%%%%%%%%%%%%%%%%%%%%%
\begin{figure}[!htb]
  \begin{center}
\includegraphics[width= 10 cm]{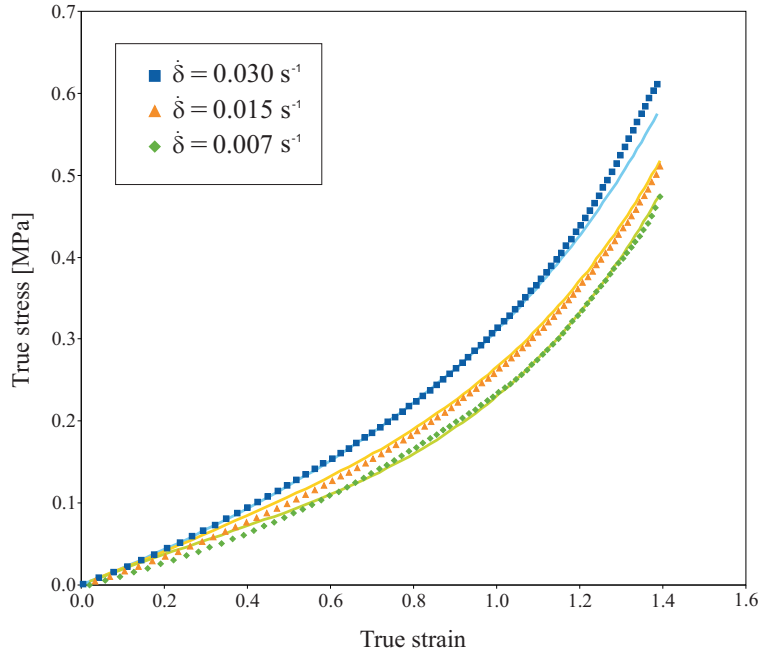}
\vspace{0.5 cm} \caption{{Viscoelastic behaviour of VHB-4910: stress response at different strain rates as obtained from parameter identification. Dots: experimental data based on experiments by Tagarielli et al.\ \cite{Tagarielli}; solid lines: simulated data.}}
    \lb{FitMechPar}
 \end{center}
\end{figure}
%%%%%%%%%%%%%%%%%%%%%%%%%%%%%%%%%%%%%%%%%%%%%%%%%%%%%%%%%%%%%%%%%%%%%%

We use a simplex search method, i.e. the Nelder-Mead algorithm for numerical minimisation. Only one Maxwell element is used in the calibration, so that $\alpha = 1$. Indeed, for the experimental data considered, adding more Maxwell elements does not substantially improve the fitting. Fig.~\ref{FitMechPar} shows the comparison between simulated and experimental data. The solid lines represent the simulated data, whereas the dots correspond to the experimental data, cf.\ \cite{Tagarielli}. The obtained material parameters are shown in Tab.\ \ref{MechPar}.

The relaxation time for the Maxwell's rheological element can be computed according to the following relation
\beq\lb{RelaxationTime}
\tau=\frac{1}{\frac{1}{2} \, \beta\,  \mu \, \dot{\varGamma}}\, .
\eeq
With the calibrated material parameters, this equation renders $\tau$  approximately equal to 45 seconds. For a similar material, namely VHB-F9473PC, a relaxation time comparable with the value resulting from our calibration is found in \cite{Kovacs}.

\begin{table}[h]
\centering
\caption{{Mechanical material parameters.}}
\lb{MechPar}
\smallskip
\begin{tabular}{ccc}
\hline
\multicolumn{1}{c}{$\mu$ [\mbox{MPa}]} &
\multicolumn{1}{c}{ $\beta$} &
\multicolumn{1}{c}{ $\dot{\varGamma}$ [s$^{-1}$ MPa$^{-1}$] } \\
\hline
\hline
 0.02746 & 1.46846 &  1.10174\\
\hline
\end{tabular}
\end{table}

\subsection{Calibration of the electrical behaviour}

In order to calibrate the electrical response of the model and to assess the electrostrictive behaviour of VHB-4910, experimental data are used for the relative dielectric permittivity at different equi-biaxial stretches. In \cite{Tagarielli} two different frequencies $\bar{f}$ are considered, namely $10^{-3}$ Hz and 200 kHz.
The experimental data, see Fig.~\ref{FitElPar}, show that $\epsilon^0_{r, 10^{-3} {\rm Hz}}=6.4$ and $\epsilon^0_{r, 200 {\rm kHz}}=3.8$ and suggests to model the dependency of the relative dielectric permittivity $\epsilon_r$ on the mechanical deformation through the first invariant $I_1$ according to the following relation
\beq\lb{varPermittivity}
\epsilon_r(\lambda_1,  \lambda_2,  \lambda_3 )= \frac{A}{\alpha_0+\alpha_1 \arctan(\alpha_2+\alpha_3(I_1(\lambda_1,  \lambda_2,  \lambda_3 )-3))}\, ,
\eeq
where $A$, $\alpha_0$, $\alpha_1$, $\alpha_2$, $\alpha_3$ are dimensionless constant parameters.
The response of the model is compared to the experimental data at different stretch levels, with the aim to find the set \{$A,\alpha_0, \alpha_1, \alpha_2,\alpha_3$\} that minimises the difference. Similar to the previous case, the error is computed as the $L_2$-norm and is then minimised by using a simplex search method. Fig.~\ref{FitElPar} shows the comparison with experimental data. The solid lines represent the prediction of the model while the dots indicate the measured permittivity, cf.\ \cite{Tagarielli}. The obtained material parameters for the relative dielectric permittivity are summarised in Tab.\ \ref{ElPar}.
%%%%%%%%%%%%%%%%%%%%%%%%%%%%%%%%%%%%%%%%%%%%%%%%%%%%%%%%%%%%%%%%%%%%%%
\begin{figure}[!htb]
  \begin{center}
\includegraphics[width= 10 cm]{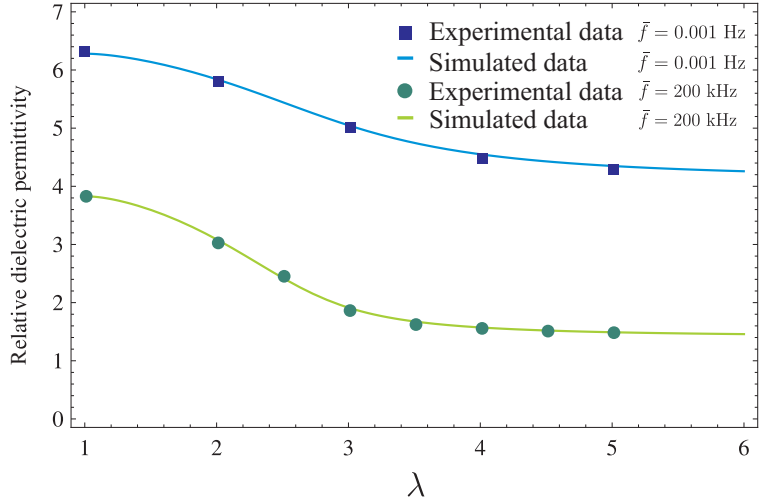}
\vspace{0.5 cm} \caption{{Dielectric permittivity of VHB-4910 at different equi-biaxial stretches for two representative frequencies $\bar f$ as obtained from parameter identification based on experiments by Tagarielli et al.\ \cite{Tagarielli}.}}
    \lb{FitElPar}
 \end{center}
\end{figure}
%%%%%%%%%%%%%%%%%%%%%%%%%%%%%%%%%%%%%%%%%%%%%%%%%%%%%%%%%%%%%%%%%%%%%%

\begin{table}[h]
\centering
\caption{{Electrical and coupling material parameters.}}
\lb{ElPar}
\smallskip
\begin{tabular}{cccccc}
\hline
 & $A$ & $\alpha_0$ & $\alpha_1$ & $\alpha_2$ & $\alpha_3$\\
\hline
\hline
 $10^{-3}$ Hz& 4.67636 & 0.85362 & $-\:$0.18891 & 0.62074 & $-\:$0.07079\\
\hline
200 kHz& 0.88568 & 0.37447 & $-\:$0.16267 & 1.20897 & $-\:$0.12075\\
\hline
\end{tabular}
\end{table}

The analysis of the DEGs presented in the next section will be based on values of the dielectric permittivity which follow the experimental data acquired at a  frequency of $10^{-3}$ Hz.

\section{Generator operating in the electrical circuit}

The performance of a soft viscoelastic dielectric elastomer generator operating in the electrical circuit, as introduced in section \ref{sec_elec},
is analysed. The dielectric elastomer material is acrylic VHB-4910 as presented above. We assume that the initial side length $L_0$ and thickness $H_0$ are equal to 100 mm and 1 mm, respectively.

We postulate that the elastomer film is initially prestretched up to a minimum value $\lambda_{min}=\lambda_o-\varLambda$, that is maintained for a sufficiently long time to allow for full relaxation. Therefore, the dielectric elastomer is connected to a source of mechanical work that stretches it periodically up to a maximum value $\lambda_{max}=\lambda_o+\varLambda$ according to the cosinusoidal relation
\beq
\lb{lambdaT}
\lambda(t)= -\varLambda \cos(\omega\, t)+ \lambda_o,
\eeq
where $\varLambda$ represents the amplitude of the stretch oscillation. In addition, $\omega=2 \, \pi f$ is the angular frequency, $f$ is the frequency of the oscillation and $\lambda_o > 1$ is the mean value of the stretch.

We solve the system of differential algebraic equations given by
the electric circuit (\ref{systemCircuit}), the nominal stress $\bS(t)$ (\ref{stress}) and the evolution equation (\ref{EvolutionLawII}) for given loading (\ref{lambdaT}) using a DAE-solver. With all relevant quantities at hand, it is possible to determine the
energies in order to evaluate the generator performance.
The input electrical energy $E_{in}$ is the integral over a cycle of the input power $P_{in}$, defined as the product of the current through the battery $i_{_{battery}}(t)$ and the voltage $\phi_o$ of the battery itself
\beq\lb{Einput} E_{in}= \int_{\rm cycle} P_{in}(t) \,\d t = \int_{\rm cycle} i_{_{battery}}(t)\,\phi_o  \,\d t \, .
\eeq
Similarly, we can calculate the total output electrical energy $E_{out}$ as the integral over a cycle of the output power $P_{out}$, defined as the product of the current through the external resistor $i_{_{load}}(t)$ and its voltage $\phi_{_{Rext}}(t)$
\beq\lb{Eoutput} E_{out}= \int_{\rm cycle} P_{out}(t)  \,\d t = \int_{\rm cycle} i_{_{load}}(t)\,\phi_{_{Rext}}(t)  \,\d t\, .
\eeq
Hence, the electrical energy produced by the generator $\Delta E=E_{out}-E_{in}$ is the difference between the electrical energy input and output. Obviously, if $\Delta E$ is positive the generator produces energy in the sense that mechanical energy is converted to electrical energy. If $\Delta E$ is negative, the generator dissipates energy, while if it is zero the generator does not convert mechanical to electrical energy.

The same net energy can be attained by subtracting the energy dissipated in the circuit ($\mathcal{D}$) from the amount of energy in the capacitor generated by the dielectric elastomer ($E_{C}$), i.e.
\beq
\Delta E= E_{out}-E_{in}= E_{C} - \mathcal{D}\, ,
\eeq
where
\beq
E_{C}=\int_{\rm cycle} P_{C}(t) \,\d t = \int_{\rm cycle} i_{C}(t) \,\phi_{C}(t) \,\d t\, .
\eeq
The energy dissipated throughout the circuit is the sum of the energy dissipated over the diode, and the two resistances $R_s$ and $R_i$, namely,
\beq
\mathcal{D}=\mathcal{D}_{D}+\mathcal{D}_{R_{s}}+\mathcal{D}_{R_{i}}\, ,
\eeq
given by
\beq\lb{CircuitryDissipation}
\begin{split}
\mathcal{D}_{D}=& \int_{\rm cycle} P_{D}(t) \,\d t = \int_{\rm cycle} i_{D}(t)\, \phi_{D}(t) \,\d t\, ,\\
\mathcal{D}_{R_{s}}=& \int_{\rm cycle} P_{R_{s}}(t) \,\d t = \int_{\rm cycle} i_{DEG}(t)\, \phi_{R_{s}}(t) \,\d t\, ,\\
\mathcal{D}_{R_{i}}=& \int_{\rm cycle} P_{R_{i}}(t) \,\d t = \int_{\rm cycle} i_{R_{i}}(t)\, \phi_{R_{i}}(t) \,\d t\, .
\end{split}
\eeq

The mechanical work performed by periodically stretching the dielectric elastomer can be determined as
\beq\lb{Wmech}
\begin{split}
W_{mech}=&   \int_{\rm cycle} \left[ S_{1}(t) \, L_0 \, H_0 \, \dot{X}_1(t) +S_{2}(t) \, L_0 \, H_0 \, \dot{X}_2(t) \right]\d t \\
=& \int_{\rm cycle} \left[ S_{1}(t) \, L_0^2 \, H_0 \, \dot{\lambda}_1(t) +S_{2}(t) \, L_0^2 \, H_0 \, \dot{\lambda}_2(t) \right]\d t \, ,
\end{split}
\eeq
where the notation $S_{i}$ is used to indicate the normal component $S_{ii}$ of the stress tensor, as depicted in Fig.~\ref{CircuitDEG}.

A measure of the performance of the generator is given by the efficiency $\eta$, defined as the ratio of the electrical energy produced by the generator and the total input energy invested. The latter is computed as the sum of mechanical work and electrical input energy,
\beq\lb{efficiency}
\eta=\frac{\Delta E}{E_{in}+W_{mech}}\, .
\eeq
For different values of the characteristic parameters of the oscillation ($\lambda_o$, $\varLambda$), we analyse the performance of the generator by varying the excitation frequency $f$ in the range {\bf from 0.1 Hz to 10 Hz}, and, as previously mentioned, the resistance of the external resistor $R_{ext}$ in the range from 0.001 G$\Omega$ to 1000 G$\Omega$.
{\bf Regarding the former range, we notice that having disregarded the inertia effects will not affect the outcome of the investigation, as an estimate of the kinetic energy involved in the motion
reveals that its maximum value in the more severe case ($f$=10 Hz, $\lambda_o=3$, $\lambda=0.5$) is only about $5\times 10^{-3}$ the amount of change of elastic strain energy stored in the material along the oscillations}.

As the relaxation time is approximately 45 seconds, see section~\ref{sec_calib}, the generator efficiency $\eta$ is computed for one cycle
after 200 seconds from the beginning of the stretch oscillation. In this context the viscous effects can be considered to be fully stabilised.

In the analysis, we compare the behaviour of the generator modelled with three constitutive responses:
\begin{enumerate}
\item hyperelastic (HYP), with constant dielectric permittivity: the energy corresponds to (\ref{WFE0}) without the viscous part and with $\epsilon^0_r=6.4$.

\item viscoelastic, with constant dielectric permittivity (VC): the energy refers to (\ref{WFE0}) with $\epsilon^0_r=6.4$.

\item viscoelastic, with electrostriction (VE): the energy is determined by (\ref{WFE0}), with deformation-dependent permittivity $\epsilon_r(\lambda_1,\lambda_2,\lambda_3)$ as discussed in eq.~(\ref{varPermittivity}).

\end{enumerate}

In the following the performance of the generator is evaluated for different loading conditions.

\subsection{Equi-biaxial loading}

We assume that the generator is subjected to equi-biaxial loading in the $\be_1$- and $\be_2$-directions, i.e.\ $S_3 = 0$. Imposing the incompressibility constraint,  the principal stretches are $\lambda_1(t)=\lambda_2(t)=\lambda(t)$ and $\lambda_3(t)=1/\lambda^2(t)$ with $\lambda(t)$ given by eq.~(\ref{lambdaT}). Hence, the deformation gradient tensor
becomes $\bF= \lambda(t)\, [\bI - \boldsymbol{e}_3\otimes\boldsymbol{e}_3] + \lambda^{-2}(t)\,\boldsymbol{e}_3\otimes\boldsymbol{e}_3 $. In this case the capacitance, as defined in (\ref{capacitance}), takes the following form
\beq\lb{EquibiCapacitance}
C=\epsilon  \, \frac{L_0^2}{H_0} \,  \lambda^4(t)
\eeq
and is thus proportional to the fourth power of the stretch.

Bearing in mind that $\boldsymbol{E}^0 = E^0(t) \,\boldsymbol{e}_3 $, with $E^0(t)=\phi{_{C}}(t)/H_0$, and using (\ref{stress}) and (\ref{displacement}), we can write the nominal electric displacement and the nominal stress in the loading directions as
\beq\lb{EquibiElDisp}
D^0(t)= \epsilon  \, \frac{\phi_{{C}}(t)}{H_0}\,   \lambda^4(t),
\eeq
\beq\lb{EquibiStress}
S_{1}(t)=S_{2}(t)=\mu \left[\lambda(t)-\frac{1}{\lambda^5(t)}\right]+\beta \,  \mu \left[\frac{\lambda(t)}{\lambda^2_v(t)}- \frac{\lambda^4_v(t)}{\lambda^5(t)}\right]-\epsilon \,  \frac{\phi^2_{{C}}(t)}{H_0^2}\ \lambda^3(t)\, .
\eeq

The internal variable $\lambda_v(t)$, with
$\boldsymbol{C}_v(t)= \lambda^2_v(t)\, [\boldsymbol{I} - \boldsymbol{e}_3\otimes\boldsymbol{e}_3] + \lambda^{-4}_v(t)\,\boldsymbol{e}_3\otimes\boldsymbol{e}_3$, is computed for the case $\alpha = 1$ and by using (\ref{EvolutionLawII}) which results in the differential equation
\beq\lb{DiffEqLambdaVisco}
\dot{\lambda}_v(t)=2\,  \dot{\varGamma}\,  \beta\,  \mu \, \lambda_v(t) \left[\frac{\lambda^2(t)}{2\, \lambda^2_v(t)} - \frac{1}{3} \left[ \frac{{\lambda^2(t)}}{\lambda^2_v(t)}+\frac{\lambda^2_v(t)}{2\,  \lambda^2(t)}\right] \right]
\eeq
with the initial condition $\lambda_v(0)=\lambda_{min}$.

% ------------------------------------------------------------------------------------
\subsubsection{Cycle characterisation of a viscoelastic DEG}
% ------------------------------------------------------------------------------------

The evolution with time of the mechanical and electrical quantities of the generator is best captured by plotting, for one loading cycle, conjugated quantities like stretch $\lambda$ vs nominal stress $S$ and charge $Q$ vs voltage $\phi_{_C}+\phi_{R_s}$.
These are illustrated in Figs.~\ref{Cycle_f01Hz} and \ref{Cycle_f1Hz} for two different frequencies, i.e.\ $f=0.1$~Hz and $f = 1$~Hz, for a
viscoelastic material following model VC, assuming a prestretch $\lambda_o=3.0$, $\varLambda=0.50$ and $R_{ext}=0.1$ G$\Omega$.

In Figs. \ref{Cycle_f01Hz}.a and \ref{Cycle_f1Hz}.a cycles starting at different times $t_i = 10, 50, 100$ and $200$~s are sketched in the $\lambda-S$ diagram. The times $t_i$ are computed relative to the full-charge of the battery occurring at $0.5\, T$.
The viscous behaviour causes a perceptible hysteresis with a stabilisation occurring after almost 200 seconds.
The downward shifting of the stress is also highlighted by the crossing point in the first depicted cycle in Fig.~\ref{Cycle_f01Hz}.a, starting at $t_i=10$ s. This crossing point results from the fact that,
under cyclic loading, the resulting nominal stress $S$ is not
periodical at the beginning of the loading until the above mentioned stabilisation occurs.
In contrast, the electrical quantities, see Figs.~\ref{Cycle_f01Hz}.b and \ref{Cycle_f1Hz}.b, show almost no change over the number of loading cycles.

%%%%%%%%%%%%%%%%%%%%%%%%%%%%%%%%%%%%%%%%%%%%%%%%%%%%%%%%%%%%%%%%%%%%%%
\begin{figure}[!t]
  \begin{center}
\includegraphics[width= 7 cm]{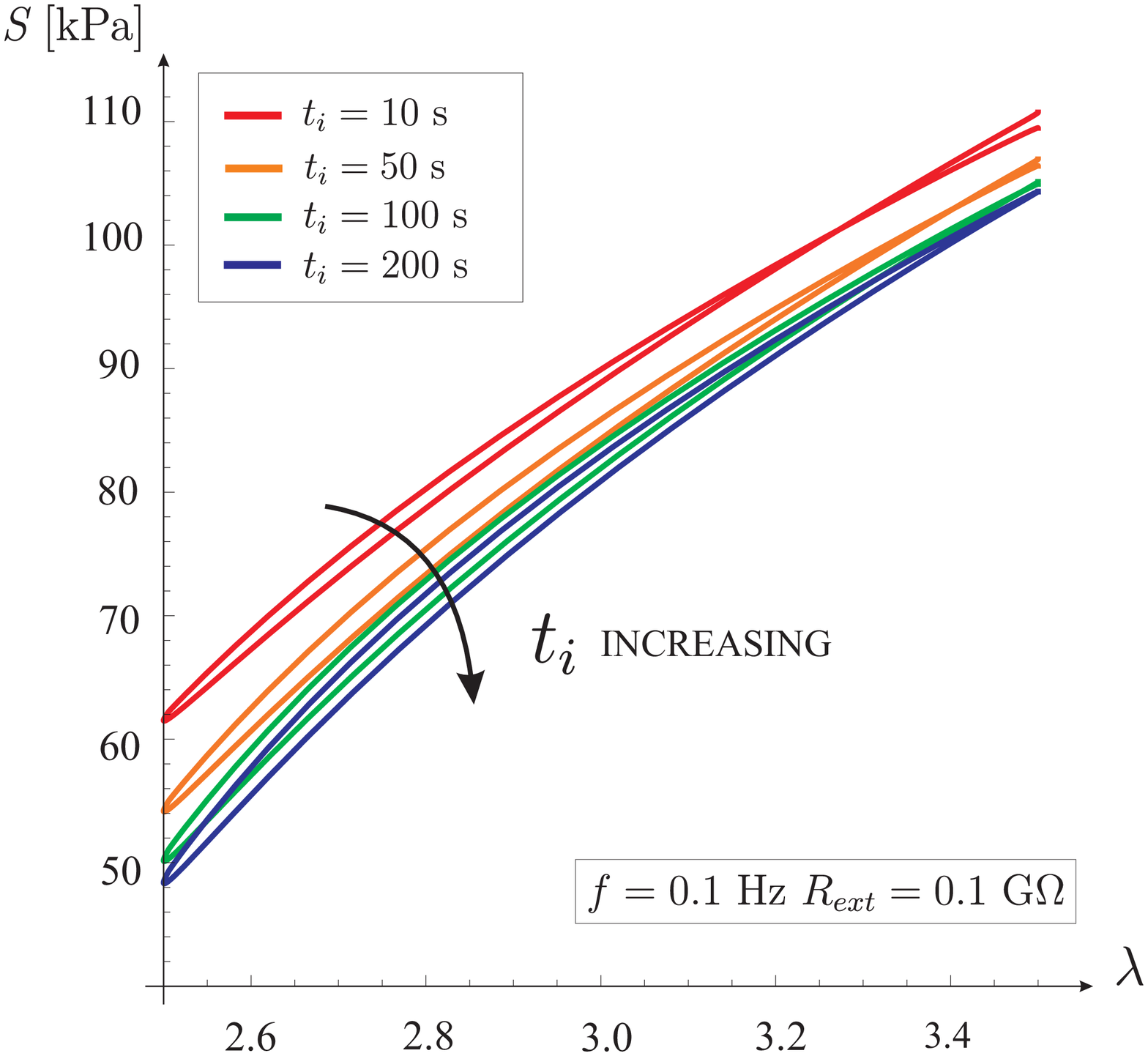}~a)
   \quad
\includegraphics[width= 7 cm]{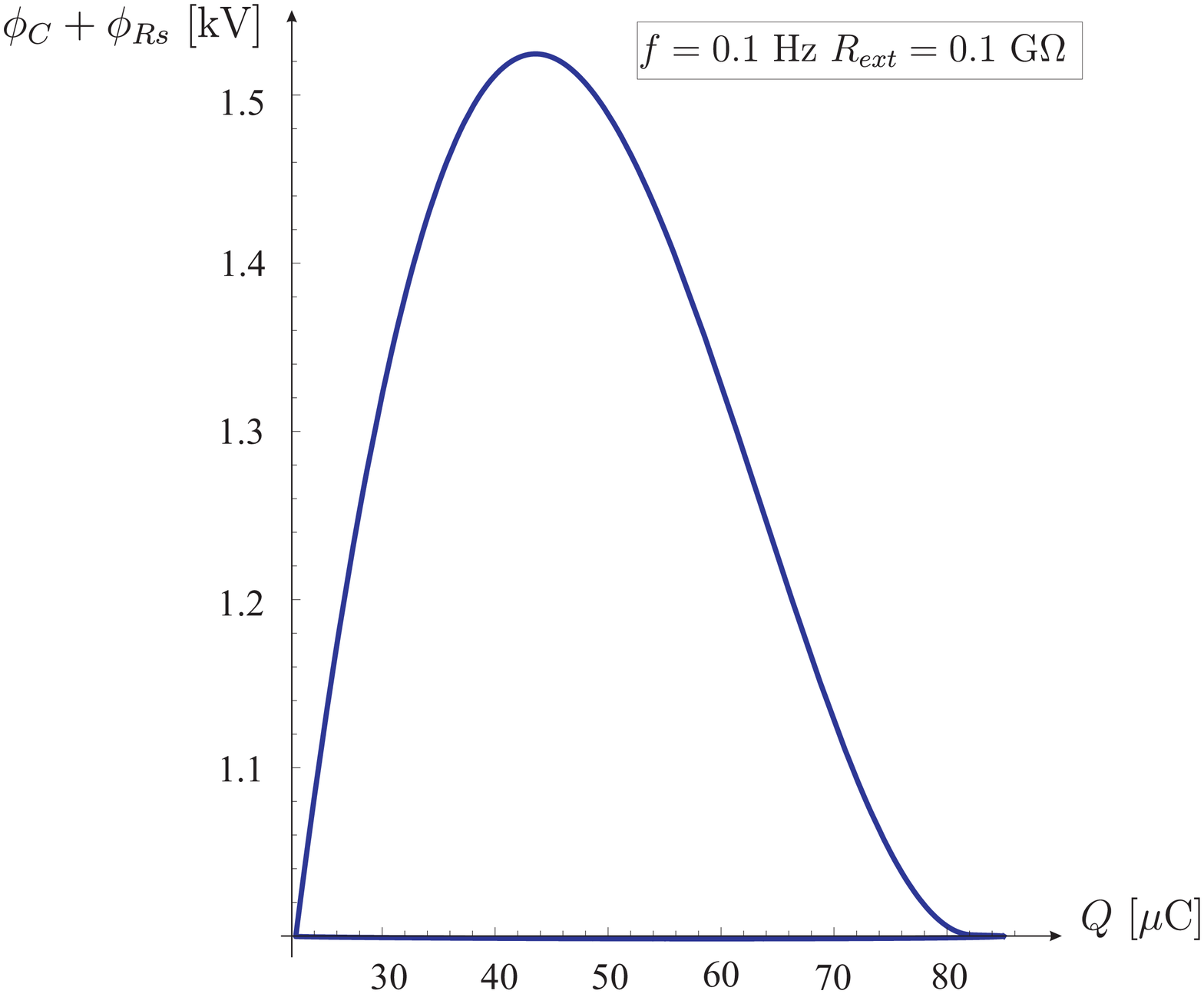}~b)
 \caption{Plot of loading cycles of a DEG a) in the mechanical and b) in the electrical planes at different initial times $t_i$, namely 10, 50, 100 and 200 seconds. Model VC, $\lambda_o=3.0$, $\varLambda=0.50$, $f=0.1$ Hz, $R_{ext}=0.1$ G$\Omega$.}
    \lb{Cycle_f01Hz}
 \end{center}
\end{figure}
%%%%%%%%%%%%%%%%%%%%%%%%%%%%%%%%%%%%%%%%%%%%%%%%%%%%%%%%%%%%%%%%%%%%%%

%%%%%%%%%%%%%%%%%%%%%%%%%%%%%%%%%%%%%%%%%%%%%%%%%%%%%%%%%%%%%%%%%%%%%%
\begin{figure}[!h]
  \begin{center}
\includegraphics[width= 7 cm]{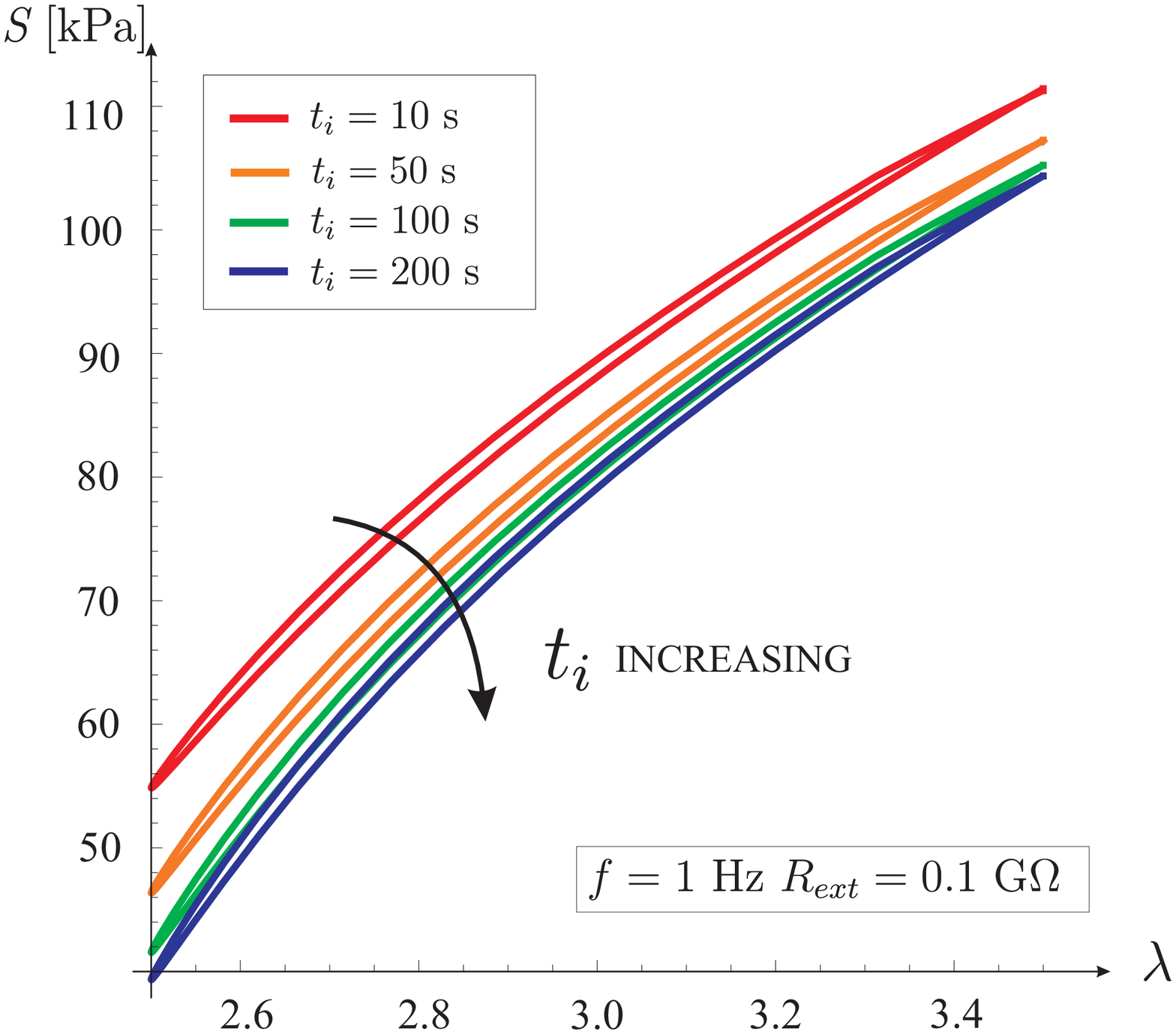}~a)
   \quad
\includegraphics[width= 7 cm]{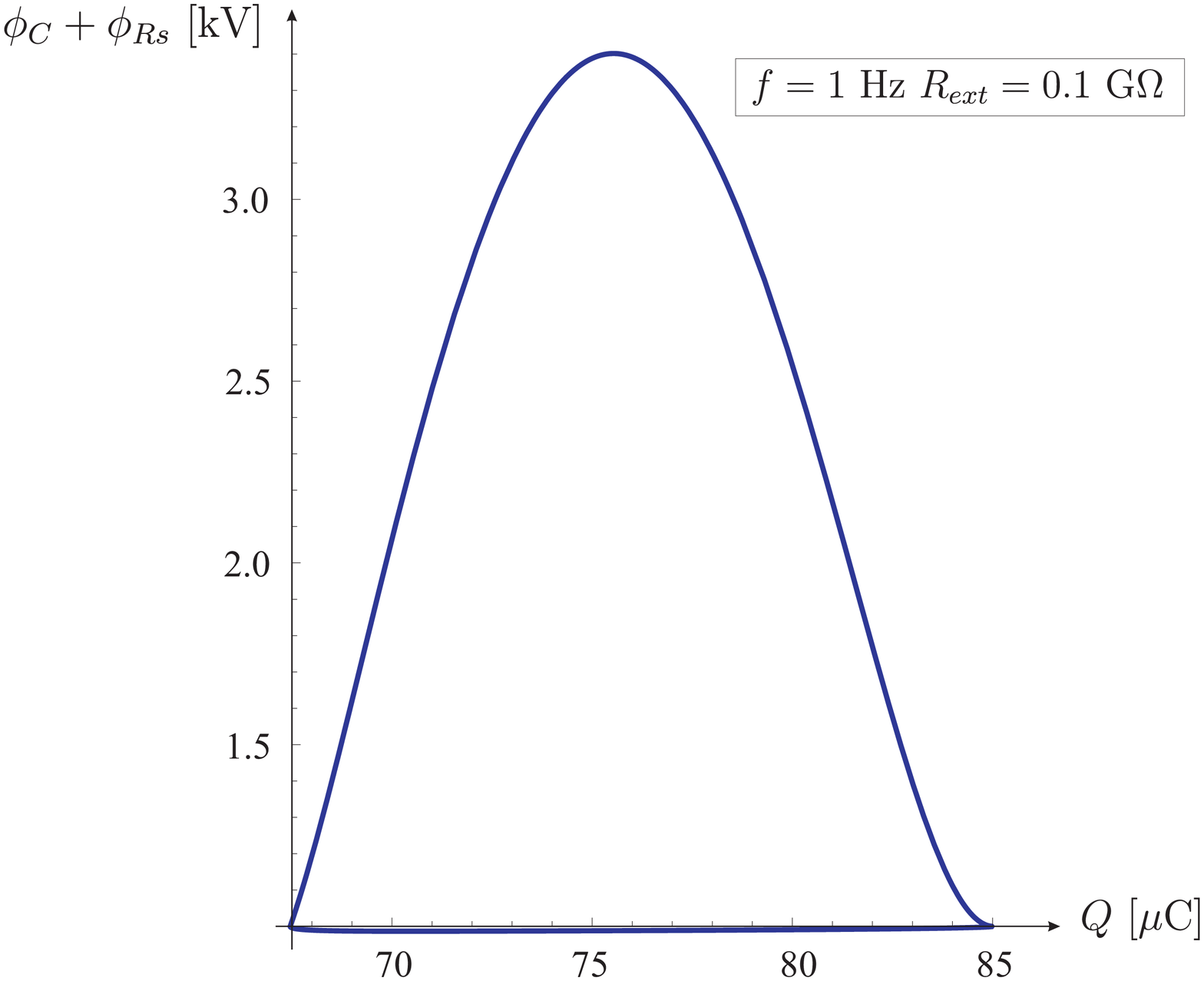}~b)
\caption{Plot of loading cycles of a DEG a) in the mechanical and b) in the electrical planes at different initial times $t_i$, namely 10, 50, 100 and 200 seconds. Model VC, $\lambda_o=3.0$, $\varLambda=0.50$, $f=1$ Hz, $R_{ext}=0.1$ G$\Omega$.}
    \lb{Cycle_f1Hz}
 \end{center}
\end{figure}
%%%%%%%%%%%%%%%%%%%%%%%%%%%%%%%%%%%%%%%%%%%%%%%%%%%%%%%%%%%%%%%%%%%%%%

%%%%%%%%%%%%%%%%%%%%%%%%%%%%%%%%%%%%%%%%%%%%%%%%%%%%%%%%%%%%%%%%%%%%%%
\begin{figure}[h]
  \begin{center}
\includegraphics[width= 11 cm]{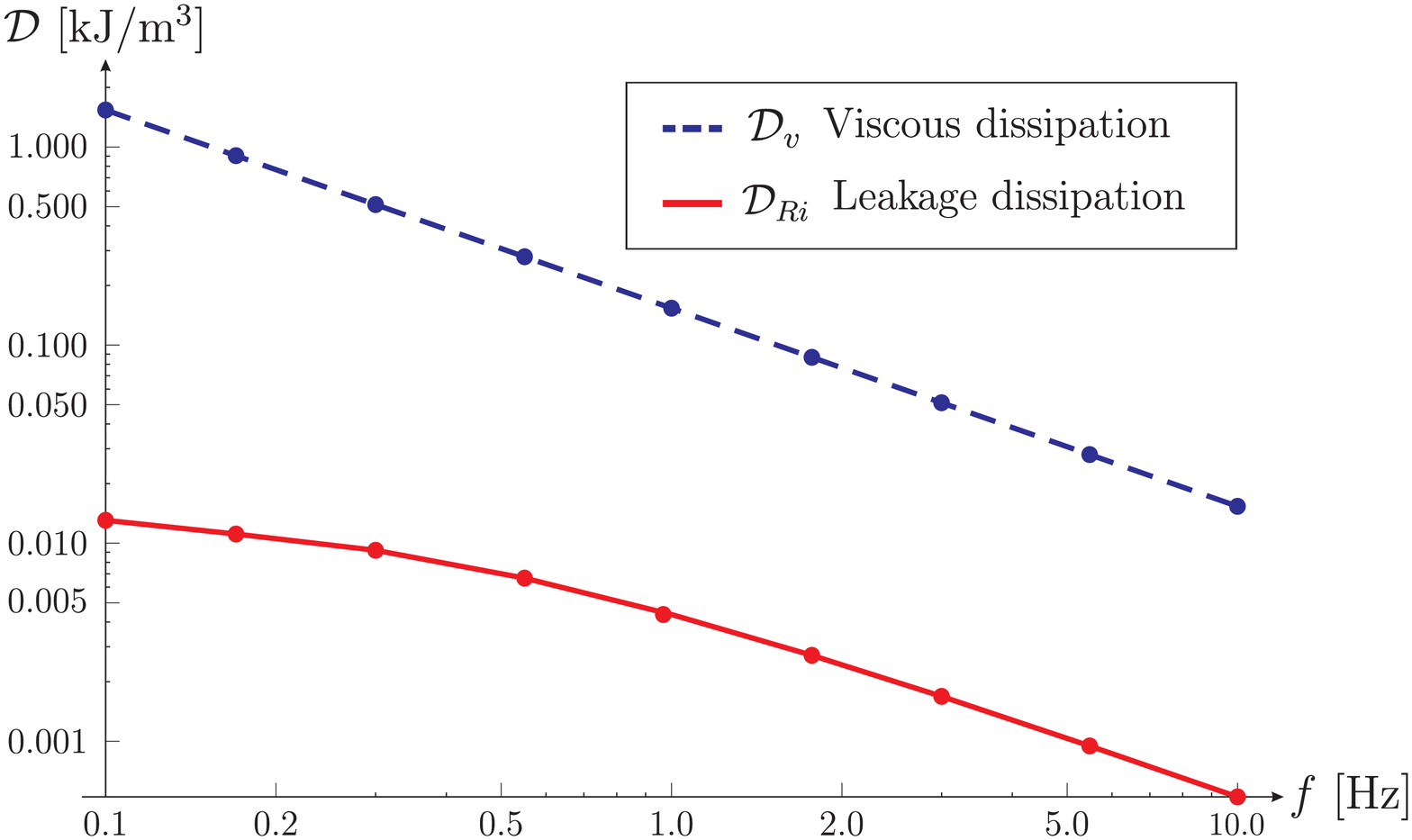}
\caption{Plot of the viscous dissipation $\mathcal{D}_v$ and the leakage dissipation $\mathcal{D}_{R_i}$ at different frequencies $f$. Model VC, $\lambda_o=3.0$, $\varLambda=0.50$, $R_{ext}=0.1$ G$\Omega$.}
    \lb{ViscousLeakageDissipation}
 \end{center}
\end{figure}
%%%%%%%%%%%%%%%%%%%%%%%%%%%%%%%%%%%%%%%%%%%%%%%%%%%%%%%%%%%%%%%%%%%%%%

The analysis of the dissipation in the generator is depicted in
Fig.~\ref{ViscousLeakageDissipation}. We computed during one loading cycle at time $t=200$ s for different excitation frequencies the specific
viscous dissipation $\mathcal{D}_v$ and the dissipation $\mathcal{D}_{R_i}$ due to the leakage current $i_{R_i}$.
Contrary to \cite{FooKohKeplinger}, {\bf and due to the low voltage applied to the circuit}, we observe that dissipation due to viscosity is always dominant in comparison to the dissipation resulting from the leakage current in the investigated range of frequencies.

In view of the energy performance of the investigated DEGs, Tab. \ref{CyclePerformance_Hyperelasticity} summarises the net energy, the mechanical work and the efficiency. All values are computed for one load cycle at $t=200$ s.
We note that the net converted energy turns out to be identical for HYP and
VC models as, for both, the electric permittivity is independent of the stretch, even though it is necessary for the viscoelastic DEG to carry out more mechanical work.
Clearly, the VE model predicts a strong reduction in the produced energy due to the decrease of the permittivity with the stretch.
More specific comments on the efficiency $\eta$ are made in section \ref{effeff}.

\begin{table}[!h]
\caption{Energy produced by the generator and mechanical work invested at two different frequencies, $f=0.1$ Hz and $f=1$ Hz, computed after 200 s for
the three material models considered: $\lambda_o=3$, $\varLambda=0.50$, $\epsilon_r^0=6.4$, $R_{ext}=0.1$ G$\Omega$. The reference volume $V_0$ is given by $L_0^2\, H_0$.}
\lb{CyclePerformance_Hyperelasticity}
\medskip
\centering
 $\lambda_o=3.0$,\ $\varLambda = 0.5$,\  $R_{ext}=0.1$ G$\Omega$  \\
\begin{tabular}{ccccc}
\hline
 & $f$ [Hz]   & $\Delta E/V_0$ [kJ/m$^3$] & $W_{mech}/V_0$ [kJ/m$^3$]& $\eta$ \\
\hline
\hline
HYP &0.1 & 1.763 & 1.792  &  13.48 \% \\
\cline{2-5}
& 1.0  & 2.456 & 2.482 &  55.95 \% \\
\hline
\hline
VC & 0.1  & 1.763 & 3.419 & 11.99 \% \\
\cline{2-5}
& 1.0  & 2.456 & 2.645 & 53.94 \% \\
\hline
\hline
VE & 0.1  & 0.374 & 2.068 & 3.01 \% \\
\cline{2-5}
& 1.0  & 1.661 & 2.032 & 44.32 \% \\
\hline
\end{tabular}
\end{table}

We close this subsection with a comment on the maximum admissible amplitude of the oscillation $\varLambda$. Once an initial prestretch is applied, followed by an in-plane tensile stress imposed in the dielectric elastomer film, a sufficient requirement along the cosinusoidal cycles is that the stress should always remain positive at any time of the loading history in order to avoid any kind of buckling or wrinkling instability.
For a hyperelastic formulation, this is achieved by simply controlling that $\lambda>1$, whereas, for a viscoelastic material, the maximum amplitude $\varLambda_{max}$ must be computed carefully for the selected material, depending on the mean stretch $\lambda_o$ and the excitation frequency. For VHB-4910 a numerical estimation is reported in Tab.\ \ref{MaxLambda} for $R_{ext}=0.1$ G$\Omega$ using model VC.
At a given $\lambda_o$, the corresponding $\varLambda_{max}$ was obtained by letting the system oscillate until stabilisation of the cycle
and then taking the value at which $\displaystyle \min_{t}\{ {S_i(t)}\}\approx 0$.
We observe that this relation is independent of the frequency, whereas
$\varLambda_{max}$ depends on the external electric resistance.
The values summarised in Tab.\ \ref{MaxLambda} clearly show the influence of viscoelasticity on the limitation of the admissible oscillation width.

\begin{table}[!h]
\caption{Maximal oscillation amplitude $\varLambda_{max}$ achievable in an equi-biaxial test
without inducing in-plane negative stresses. Model VC, $R_{ext}=0.1$ G$\Omega$.}\lb{MaxLambda}
\medskip
\centering
\begin{tabular}{lcccc}
\hline
 $\lambda_o$     & 1.8  & 2.0  & 3.0  & 4.0 \\
 $\varLambda_{max}$ & 0.30 & 0.38 & 0.69 & 0.88 \\
\hline
\end{tabular}
\end{table}

\subsubsection{Efficiency analysis}

\lb{effeff}

The generator efficiency $\eta$ calculated by means of (\ref{efficiency}) and by using (\ref{Wmech}) and (\ref{EquibiStress}) is now investigated
in terms of the imposed frequency and the external electrical resistance. Plots of $\eta(f,R_{ext})$ for the three considered constitutive models
and $\lambda_o = 3$ are shown in Fig.~\ref{Efficiency_EquiBiaxial_3Dplot}. Three amplitudes $\varLambda$
are analysed in every chart, namely $\varLambda=0.50$, $\varLambda=0.25$ and $\varLambda=0.10$. The frequency is examined up to 10 Hz, even though the maximum operational frequency for DEG devices of the type analysed here is usually in the order of a few Hz.

%%%%%%%%%%%%%%%%%%%%%%%%%%%%%%%%%%%%%%%%%%%%%%%%%%%%%%%%%%%%%%%%%%%%%%
\begin{figure}[!htb]
  \begin{center}
\includegraphics[width= 8.5 cm]{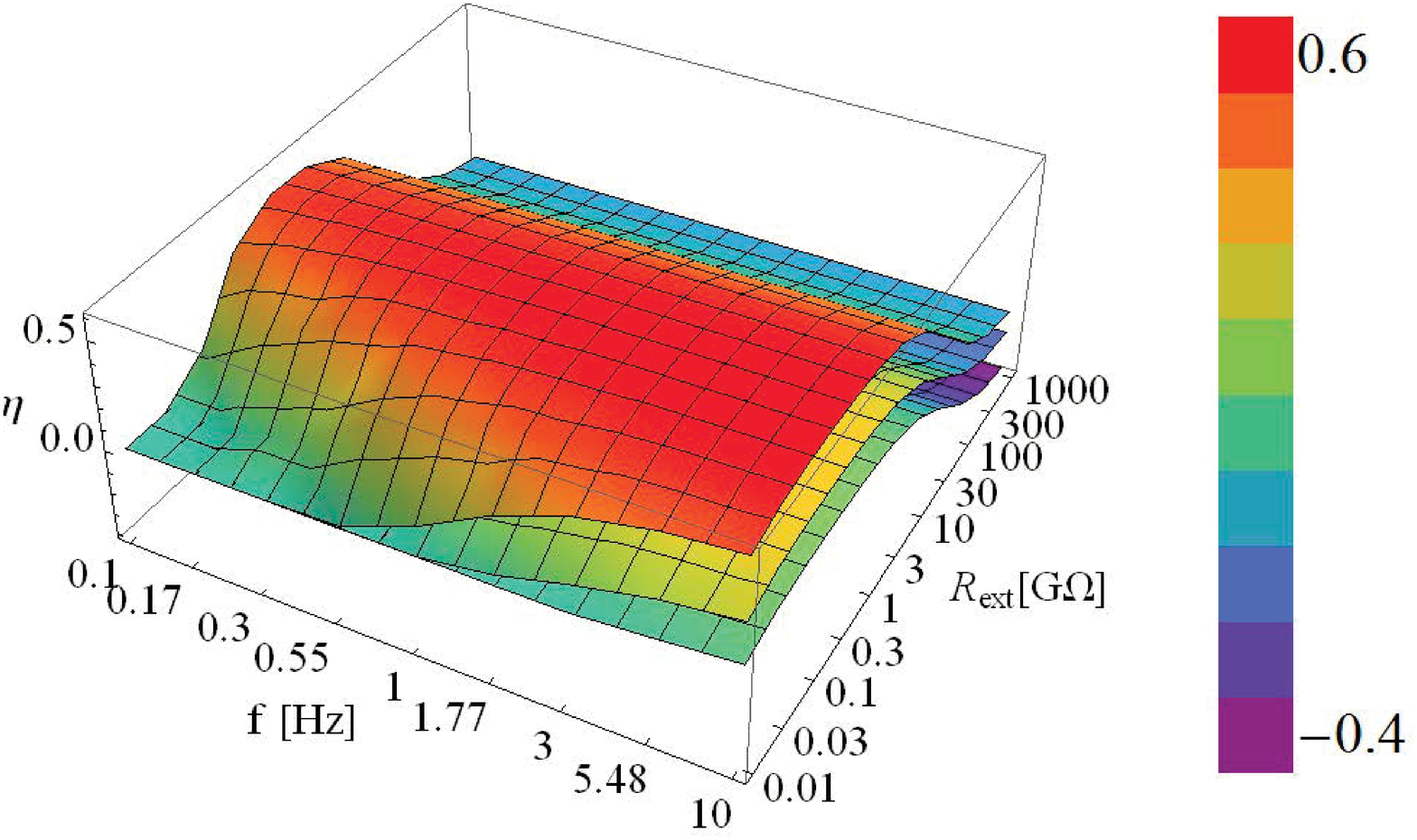}~a)
\includegraphics[width= 8.5 cm]{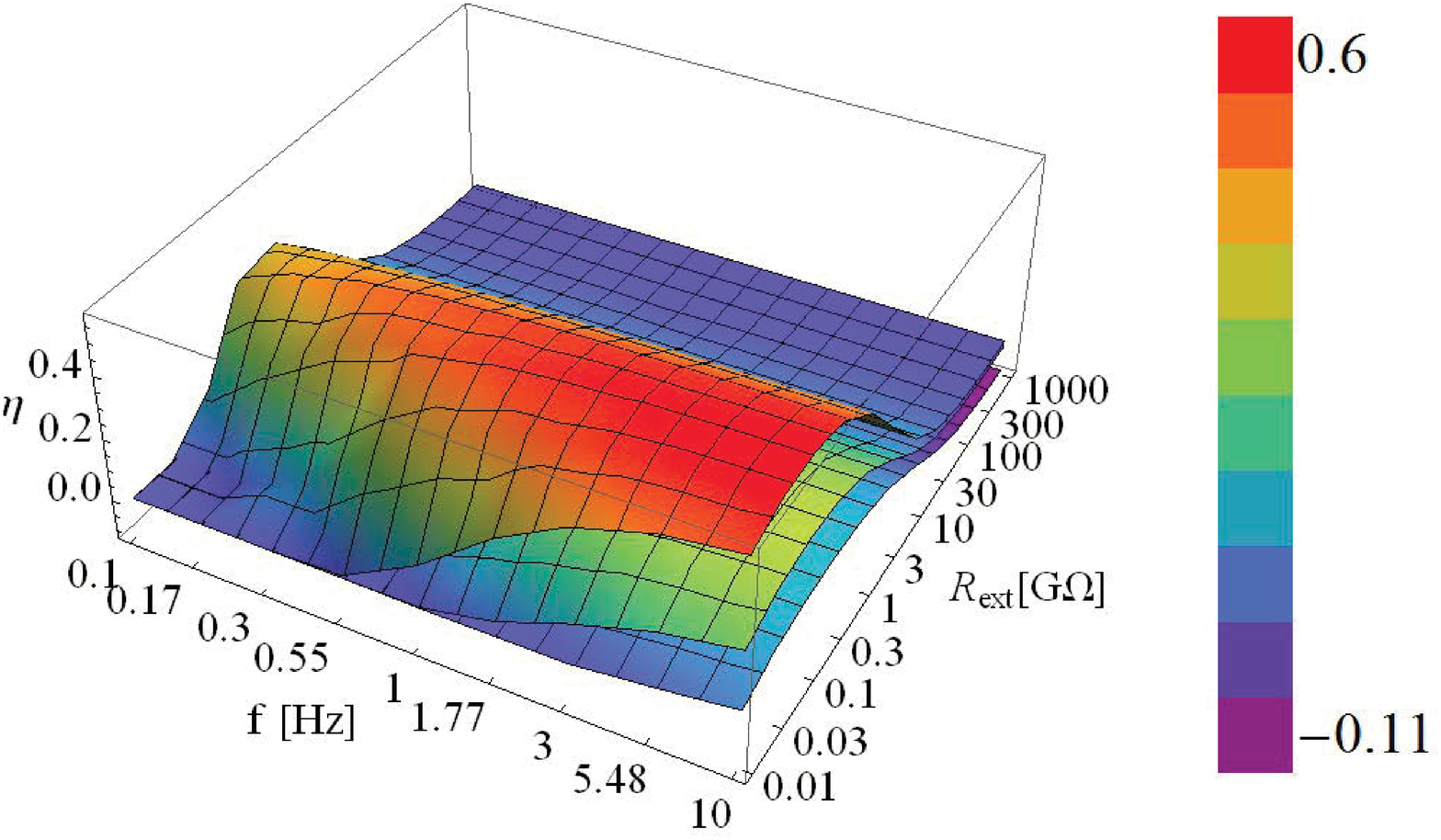}~b)
\includegraphics[width= 8.5 cm]{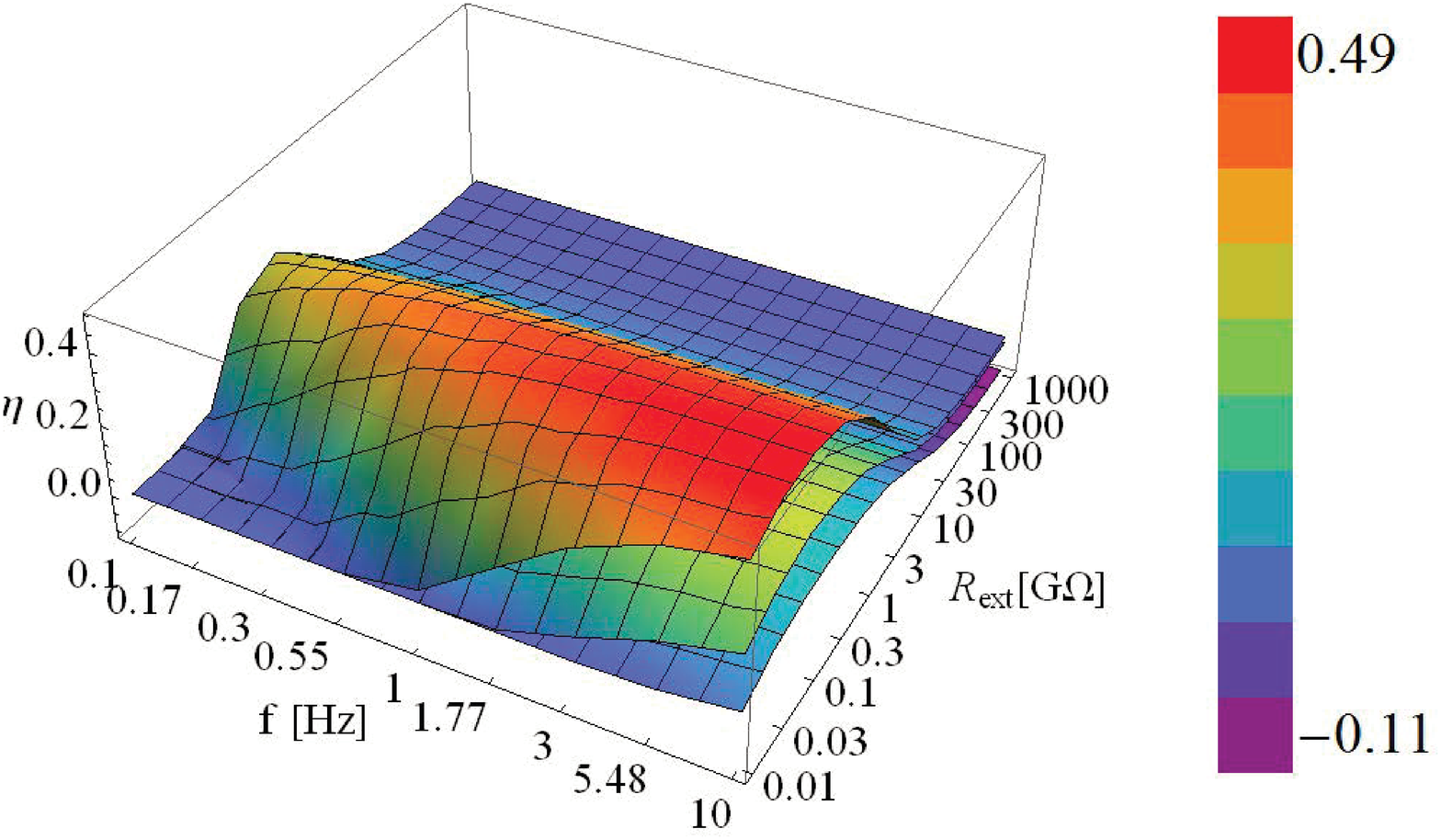}~c)
\caption{{Plot of the efficiency $\eta (R_{ext},f)$ for the three different material models:
a) hyperelastic, HYP, b) viscoelastic, VC, and c) electrostrictive viscoelastic, VE.
Equi-biaxial loading conditions with $\lambda_o=3.0$; $\varLambda=0.50$, $\varLambda=0.25$ and $\varLambda=0.10$.}}
    \lb{Efficiency_EquiBiaxial_3Dplot}
 \end{center}
\end{figure}
%%%%%%%%%%%%%%%%%%%%%%%%%%%%%%%%%%%%%%%%%%%%%%%%%%%%%%%%%%%%%%%%%%%%%%

Firstly, we note that the efficiency $\eta$ could be either positive or negative depending on the values of the external resistance $R_{ext}$.
Negative values for $\eta$ are observed for $R_{ext}$ taking values greater than 30 G$\Omega$ in the case of small oscillation amplitudes $\varLambda$.
An evident outcome of the data is that the hyperelastic (HYP) model always predicts higher efficiency in comparison to both viscoelastic models.
Moreover, larger amplitudes are always associated with larger efficiency, irrespective of the material model.
The reason for this is that the capacitance of the generator depends on the stretch to the power of four which results in considerable increase of the output electrical energy.
On the contrary, the energy supplied to the system shows a less than proportional increase in the oscillation amplitude $\varLambda$.

Tab.\ \ref{CyclePerformance_amplitude} shows these energy figures for the three selected amplitudes.
In addition, we observe that the difference between the three material models is more pronounced for high values of $\varLambda$,
as shown in Figs. \ref{Efficiency_EquiBiaxial_2Dplot}.a and  \ref{Efficiency_EquiBiaxial_2Dplot}.b.

\begin{table}[!h]
\caption{Energy produced by the generator and mechanical work invested for the three selected amplitudes $\varLambda=0.10$, $\varLambda=0.25$ and $\varLambda=0.50$, computed after 200 s for the VC model: $\lambda_o=3$, $f=1$ Hz, $\epsilon_r^0=6.4$, $R_{ext}=0.1$ G$\Omega$. The reference volume $V_0$ is given by $L_0^2\,  H_0$.}
\lb{CyclePerformance_amplitude}
\medskip
\centering
$\lambda_o=3.0$,\ $f = 1$ Hz,\  $R_{ext}=0.1$ G$\Omega$,  VC model  \\
\begin{tabular}{cccccc}
\hline
  $\varLambda$   &$E_{in}/V_0$ [kJ/m$^3$] &$E_{out}/V_0$ [kJ/m$^3$] & $\Delta E/V_0$ [kJ/m$^3$] & $W_{mech}/V_0$ [kJ/m$^3$] & $\eta$\\
\hline
\hline
 0.10 & 1.067 & 1.142 & 0.075 & 0.085  &  6.49 \% \\
\hline
 0.25 & 1.303 & 1.771 & 0.468 & 0.516  & 25.74 \%\\
\hline
 0.50 & 1.906 & 4.362 & 2.465 & 2.645 & 53.94 \% \\
\hline
\end{tabular}
\end{table}

Fig.~\ref{Efficiency_EquiBiaxial_2Dplot}.a displays the efficiency comparison for the three constitutive models in the case of $\lambda_o = 3$ and $R_{ext}= 1\, \mbox{G}\Omega$, as data show that the highest efficiency values lie close to this value, cf. Fig.~\ref{Efficiency_EquiBiaxial_3Dplot}.
%It is evident that for small values of the oscillation amplitude $\varLambda$ the results given by the three different models are almost the same, while when the oscillation amplitude $\varLambda$ increases the results differs more and more.
{\bf For $\varLambda=0.5$ the efficiency difference between models HYP and VC is around $15\%$ while that between HYP and VE is approximately $23\%$. This difference reduces respectively  to $5.3\%$ and $9.5\%$ for $\varLambda=0.25$, and to $0.6\%$ and $2.4\%$ for $\varLambda=0.1$.
{\bf The stretch dependency of the permittivity accounted in model VE reduces $\eta$ to approximately 8\%  (2\%)}
with respect to the efficiency of the classical electro-viscoelastic model VC for $\varLambda=0.5$ ($\varLambda=0.1$).}

The same comparison for $\lambda_o = 3$ and $f = 1\ \mbox{Hz}$ in terms of the external resistance $R_{ext}$
is depicted in Fig.~\ref{Efficiency_EquiBiaxial_2Dplot}.b. As already observed, $\eta$ is negative for high values of the external resistance $R_{ext}$, depending on the value of the oscillation amplitude $\varLambda$, in the range between 30 and 300 G$\Omega$ (increasing values for increasing $\varLambda$'s).
%For small value of the oscillation amplitude, i.e. $\varLambda=0.1$, this occurs for $R_{ext}> 30$ G$\Omega$, while for higher value of the amplitude the efficiency is negative for higher values of the external resistance $R_{ext}$, namely 100 G$\Omega$ for $\varLambda=0.25$ and 300 G$\Omega$ $\varLambda=0.5$.

In these cases, the output electrical energy is lower than the input one.
An explanation is that the voltage of the connected battery, $\phi_o$=1 kV, is not sufficient
to power the mechanical energy conversion. As a result, the charge exchanged by the generator at
every cycle is relatively low and inadequate to feed the external resistor.
For a battery operating at a higher voltage, the threshold value of $R_{ext}$,
beyond which $\eta< 0$, increases accordingly.

Among the three models, hyperelasticity predicts a wider range where the efficiency is positive.
For small values of $R_{ext}$, the VC model behaves similarly to the hyperelastic one up to a peak value, which occurs at lower values of the external resistance $R_{ext}$ increasing the amplitude $\varLambda$.
Moreover, it is noted that, for the model with   electrostriction (VE), the values of the efficiency are always lower in comparison to the hyperelastic model within the whole considered range of $R_{ext}$.

The influence of the mean stretch $\lambda_o$ on the efficiency in terms of the external frequency $f$ is outlined in Fig.~\ref{CompEtaLo} for $R_{ext}=1$ G$\Omega$ and for a generator based on the viscoelastic (VC) constitutive assumption.
When $\lambda_o$ is equal to 1.8 the behaviour of the generator is noticeably {\bf different between frequencies lower and higher than 1 Hz}: the change in $\eta$ through the frequency range is {\bf approximately $19\%$} for $\varLambda=0.1$ raising to $33\%$ for $\varLambda=0.25$. On the contrary, for a higher mean stretch ($\lambda_o=3$), the behaviour of the generator is more stable, the efficiency variation is up to $6\%$ for the considered values of the amplitude. Hence, for a viscoelastic DEG, when the average value of the oscillation $\lambda_o$ increases, the behaviour of the generator becomes more stable and less dependent on the other electrical and mechanical
parameters.

%Hence, from the numerical data, it is evident that for large amplitudes of the oscillation it is highly important to take into account time-dependent effects, since these reduce significantly the efficiency of the generator compared with the hyperelastic modelling.

%%%%%%%%%%%%%%%%%%%%%%%%%%%%%%%%%%%%%%%%%%%%%%%%%%%%%%%%%%%%%%%%%%%%%%
\begin{figure}[!htb]
  \begin{center}
\includegraphics[width= 8 cm]{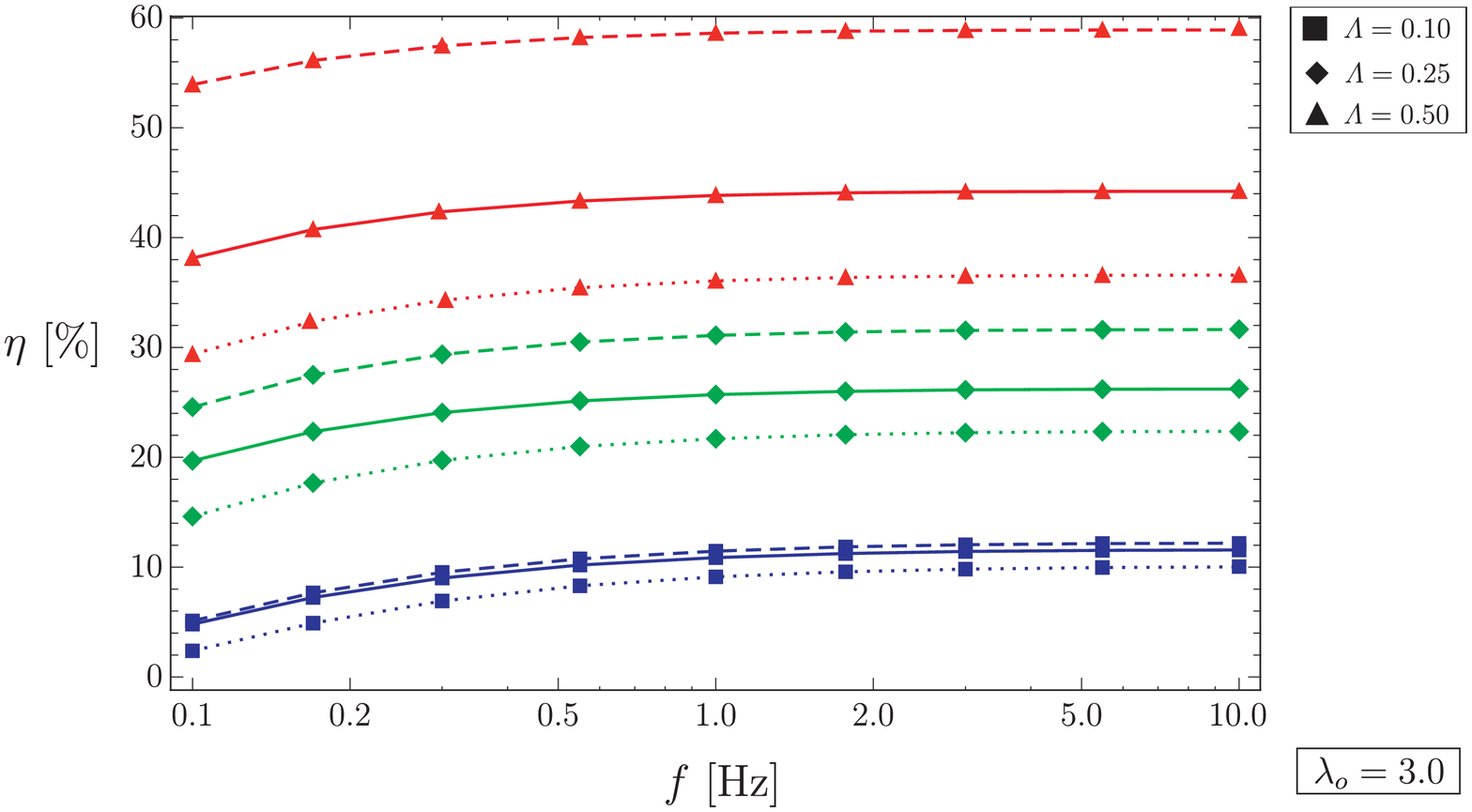}~a)
\includegraphics[width= 8 cm]{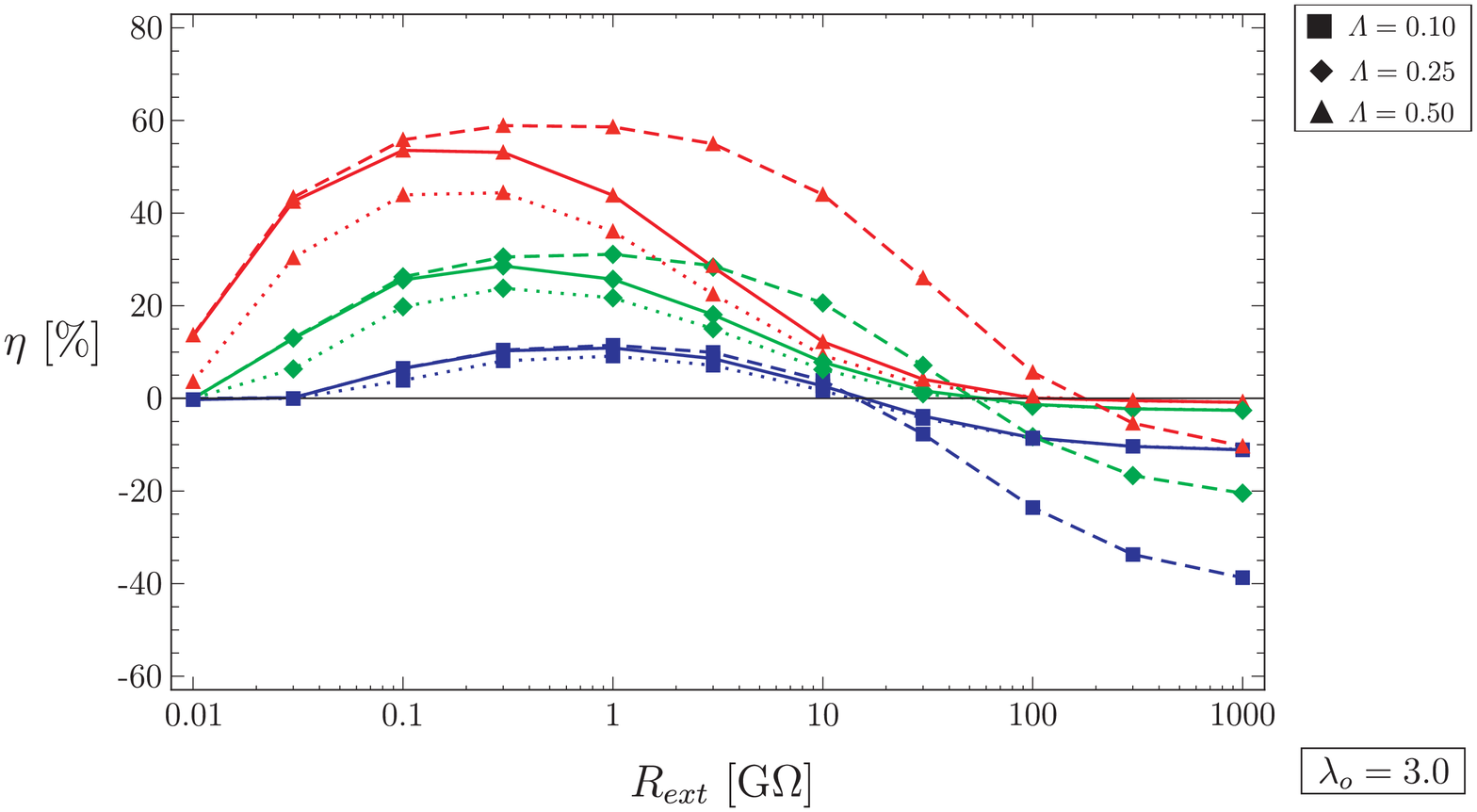}~b)
\caption{Plot of the efficiency $\eta$ versus a) frequency $f$ at $R_{ext}=$  1 G$\Omega$, and b) external resistance $R_{ext}$ at $f=1$ Hz. Equi-biaxial loading conditions with $\lambda_o=3.0$; $\varLambda=0.50,\ 0.25,\ 0.10$. Dashed, continuous and dotted lines are referred respectively to HYP, VC and VE models.}
    \lb{Efficiency_EquiBiaxial_2Dplot}
 \end{center}
\end{figure}
%%%%%%%%%%%%%%%%%%%%%%%%%%%%%%%%%%%%%%%%%%%%%%%%%%%%%%%%%%%%%%%%%%%%%%

%%%%%%%%%%%%%%%%%%%%%%%%%%%%%%%%%%%%%%%%%%%%%%%%%%%%%%%%%%%%%%%%%%%%%%
%\begin{figure}[!htb]
%  \begin{center}
%\includegraphics[width= 10 cm]{figure/CompEtaf1.eps}
%\vspace{0.5 cm} \caption{\footnotesize{Efficiency $\eta(R_{ext},f)$ for solicitation frequency $f$ equal to 1 Hz: $\lambda_o=3.0$; $\varLambda=0.50,\ 0.25,\ 0.10$. Dashed lines are referred to an hyperelastic material, continuous lines to a viscoelastic one with constant relative dielectric permittivity, dotted lines to a viscoelastic one with strain dependent relative dielectric permittivity.}}
%    \lb{CompEtaf1}
% \end{center}
%\end{figure}
%%%%%%%%%%%%%%%%%%%%%%%%%%%%%%%%%%%%%%%%%%%%%%%%%%%%%%%%%%%%%%%%%%%%%%

%%%%%%%%%%%%%%%%%%%%%%%%%%%%%%%%%%%%%%%%%%%%%%%%%%%%%%%%%%%%%%%%%%%%%%
\begin{figure}[!htb]
  \begin{center}
\includegraphics[width= 10 cm]{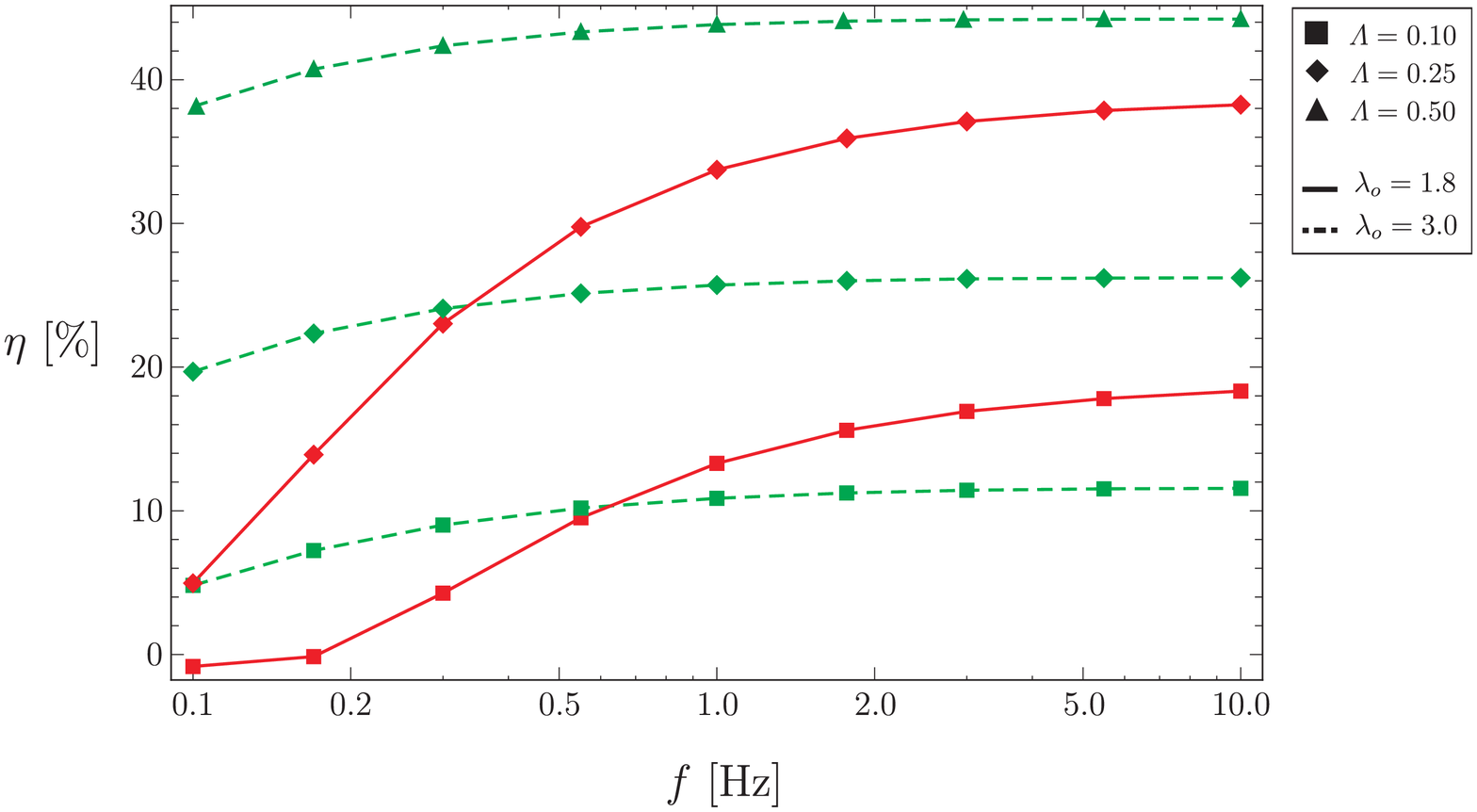}
\caption{Plot of the efficiency $\eta$ versus frequency $f$ for two values of the mean value of the oscillation
stretch $\lambda_o=1.8$ and  $\lambda_o=3$. Equi-biaxial loading conditions with $R_{ext}=$ 1 G$\Omega$, VC model.}
    \lb{CompEtaLo}
 \end{center}
\end{figure}
%%%%%%%%%%%%%%%%%%%%%%%%%%%%%%%%%%%%%%%%%%%%%%%%%%%%%%%%%%%%%%%%%%%%%%

\subsection{Uniaxial loading}

The soft dielectric elastomer here is subjected to uniaxial loading conditions in the direction $\be_1$ so that $S_2=S_3=0$. Imposing the incompressibility constraint, the principal stretches are $\lambda_1(t)=\lambda(t)$ and $\lambda_2(t)=\lambda_3(t)=1/\sqrt{\lambda(t)}$.
Hence, the deformation gradient tensor becomes $\bF= \lambda(t)\, \boldsymbol{e}_1\otimes\boldsymbol{e}_1 + 1/\sqrt{\lambda(t)}\,[\bI - \boldsymbol{e}_1\otimes\boldsymbol{e}_1] $.
Compared with the biaxial case, the capacitance is lower as it shows only a direct proportionality to the axial stretch, i.e.
\beq
\lb{UniCapacitance}
C=\epsilon  \, \frac{L_0^2}{H_0}\,   \lambda(t)\, .
\eeq

Bearing in mind that $\boldsymbol{E}^0 = E^0(t) \,\boldsymbol{e}_3 $, with $E^0(t)=\phi{_{C}}(t)/H_0$, we can write the nominal electric displacement and the nominal stress in the loading direction as
\beq\lb{UniElDisp}
D^0(t)= \epsilon \, \frac{\phi_{_{C}}(t)}{H_0} \,  \lambda(t)\, ,
\eeq
while the relation between stress, stretch and voltage turns out to be
\beq\lb{UniStressII}
S_{1}(t)=\mu \left[ \lambda(t)- \frac{1}{\lambda(t)^2}\right] + \beta \,  \mu \,  \left[\frac{\lambda(t)}{\lambda_v(t)^2} - \frac{\lambda_v(t)}{\lambda(t)^2}\right] - \epsilon \, \frac{\phi_{_{C}}(t)^2}{H_0^2} \, .
\eeq
The internal variable $\lambda_v (t)$ is computed by integrating the evolution equation (\ref{EvolutionLawII}) which, in the incompressible uniaxial case, reduces to
\beq\lb{InternalVariableUni}
\dot{\lambda}_{v}(t)= \frac{1}{4} \, \dot{\varGamma}\, \beta \,  \mu \,  \lambda_{v}(t) \left[   \frac{\lambda(t)^2}{\lambda_{v}(t)^2} - \frac{1}{3} \left[\frac{\lambda(t)^2}{\lambda_{v}(t)^2}+ 2\, \frac{\lambda_v(t)}{\lambda(t)} \right] \right]\, ,
%\dot{\lambda}_{v}(t)= \frac{1}{4} \dot{\varGamma}\beta  \mu  \lambda_{v}(t) \left[   \frac{\lambda(t)^2}{\lambda_{v}(t)^2} - \frac{1}{3} \left[\frac{\lambda(t)^2}{\lambda_{v}(t)^2}+\frac{\lambda_2(t)^2}{\lambda_{v2}(t)^2}+\frac{\lambda_3(t)^2}{\lambda_{v3}(t)^2}\right] \right],
\eeq
with the initial condition $\lambda_v(0)=\lambda_{min}$.

\begin{figure}[!htb]
  \begin{center}
\includegraphics[width= 10 cm]{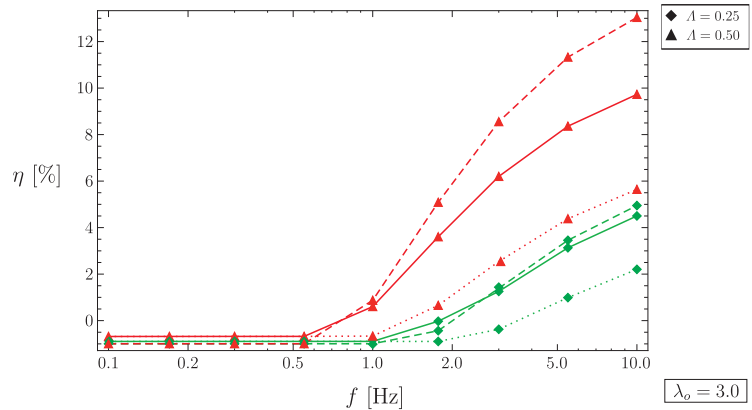}
\vspace{0.5 cm} \caption{Plot of the efficiency $\eta(R_{ext},f)$ for an external resistance $R_{ext}= $1 G$\Omega$ and $\lambda_o=3.0$, $\varLambda=0.50$ and $\varLambda=0.25$. Dashed, continuous and dotted lines are referred respectively to HYP, VC and VE models.}
    \lb{CompEtaR1u}
 \end{center}
\end{figure}
%%%%%%%%%%%%%%%%%%%%%%%%%%%%%%%%%%%%%%%%%%%%%%%%%%%%%%%%%%%%%%%%%%%%%%

%The plot of the generator efficiency $\eta (f,R_{ext})$ is shown in Fig.~\ref{Efficiency_Uniaxial_3Dplot}.
%These figures are referred to the case of an oscillation around the average value $\lambda_o = 3.0$ and decreasing value of the amplitude $\varLambda$, namely 0.50, 0.25 and 0.10, respectively in the case of hyperelastic material model and viscoelastic one, with constant and deformation dependent relative dielectric permittivity.

Three-dimensional plots of the efficiency, i.e. graphical representations of the function $\eta (f,R_{ext})$, are not given here for conciseness. But it is found that at the same supplied voltage $\phi_o$ and compared with the equi-biaxial loading, the uniaxial excitation leads to overall lower values of the efficiency. Additionally, the range of points $(f,R_{ext})$ with positive efficiency is more limited.
As in the case of equi-biaxial loading, the HYP constitutive model always predicts higher values of the efficiency with respect to the two kinds of viscoelasticities.
However, in this uniaxial loading case, the efficiency of the generator is greater than zero only for few values of the variables $f$ and $R_{ext}$. When the amplitude of the oscillation $\varLambda$ is small, i.e. $\varLambda=0.10$, the efficiency is always lower or equal to zero, i.e.\ $\eta \leq 0$, even in the case of hyperelasticity.
%For larger value of the oscillation amplitude, the efficiency gets positive but only for values of the variables in the medium-high range, that is $1 \leq f \leq 30$ Hz for the frequency and $0.1 \leq R_{ext} \leq 10 \ \mbox{G}\Omega$ for the external resistance. This is due to the fact that the dependence of the capcitance on the applied stretch is simply linearly proportional.

%For an oscillation amplitude $\varLambda = 0.5$, the efficiency difference between the HYP model and VC is around $4\%$, while between the HYP model and the VE is close to $8.5\%$. As in the previous case, these differences decrease for oscillation amplitudes $\varLambda$ decreasing.
%For VE model, the reduction in the efficiency due to electrostriction is around $5\%$, thus comparable to the case of equi-biaxial stress, decreasing with the decrease of the oscillation amplitude $\varLambda$. % For the smallest value of the oscillation amplitude, $\varLambda=0.1$ the difference reaches the same value than in the case of equi-biaxial loading (about $1\%$).

Fig.~\ref{CompEtaR1u}, obtained for $\lambda_o = 3$ and $R_{ext}=1 \mbox{G}\Omega$ with $\varLambda=0.25$ and $\varLambda=0.50$, shows
negative values of efficiency at low frequencies.
As in the case of equi-biaxial loading, the efficiency computed with the HYP model is greater than the predicted by VC and VE models.
The difference between the three different models decreases for decreasing values of the oscillation amplitude $\varLambda$.
{\bf For $\varLambda=0.5$, the difference in efficiency between HYP and VC models is approximately 1.3\% while the difference between HYP and VE models is approx.\ 3.1\%. For $\varLambda=0.25$ we obtained 0.2\% and 0.9\%, respectively.}
As mentioned before, the analysis clearly demonstrates that, by applying
the same oscillation conditions $\varLambda$ and $\lambda_o$ the uniaxial loaded generator shows a  considerably lower efficiency than the equi-biaxially loaded generator.

To relate the two loading conditions we investigate the DEG performance when the capacitance changes during a cycle are equal.
We choose the hyperelastic (HYP) model under equi-biaxial loading
$\lambda_o=1.8$ and $\varLambda=0.1$ as a reference. An equal  capacitance change is observed in a DEG subjected to the uniaxial loading for $\lambda_o=10.621$ and $\varLambda=2.34$.
The computed efficiency with $R_{ext}=1$ G$\Omega$ and $f=1$ Hz are $\eta=15.16\%$ for equi-biaxial and $\eta=13.04\%$ for uniaxial loading.

\section{Conclusions}

Soft materials usually employed in dielectric elastomer generators show a remarkable viscoelastic behaviour and may display a
deformation-dependent permittivity, a phenomenon known as electrostriction. Therefore, the design and the analysis of soft energy harvesters, which undergo a high number of electromechanical cycles at frequencies in the range of one Hertz, must be based on reliable models that include such behaviour.
In this paper, a large strain electro-viscoelastic model for a polyacrilate elastomer, VHB-4910 produced by 3M, is proposed and calibrated based on experimental data available in the literature.

The model is used to simulate the performance of a soft prestretched dielectric elastomer generator operating in a circuit where a battery at constant voltage supplies the required charge at each cycle and where an electric load consumes the produced energy. Two periodic in-plane loading conditions, namely homogeneous states under equi-biaxial and uniaxial deformation, are considered for the soft capacitor.

Application of the proposed model provides for the generator i) the assessment of viscous and electrostrictive effects in the computation of efficiency and amount of net energy gained after each cycle and ii) the evaluation of energy losses in all dissipative sources of the device as a function of the imposed mechanical frequency.

The main outcome of this analysis is that, compared with a hyperelastic model, the efficiency is reduced by viscoelasticity for high values of the mean stretch and of the amplitude of stretch oscillation. The reduction is almost insensitive of the mechanical frequency while the efficiency is further reduced by electrostrictive properties of the material. We observed a range of values of the external electric load with a maximal efficiency. Furthermore, {\bf at low applied voltage,} the viscous dissipation of the material dominates the energy loss stemming from the leakage current across the filled soft capacitor.

\section*{Acknowledgements}
The authors gratefully acknowledge Prof.\ Vito Tagarielli for providing the experimental data.
E.B.\ gratefully acknowledges support from the EU FP7 project PIAP-GA-2011-286110-INTERCER2.
M.G.\ gratefully acknowledges support from the EU FP7 project ERC-2013-ADG-340561-INSTABILITIES.

\end{document}